\documentclass[useAMS,usenatbib,usegraphicx]{mn2e}

\title[The JCMT Nearby Galaxies Legacy Survey IV. Velocity
  Dispersions in the Molecular Gas]{The 
JCMT Nearby Galaxies Legacy Survey IV. Velocity 
  Dispersions in the Molecular Interstellar Medium in Spiral Galaxies}
\author[C. D. Wilson et al.]
    {C. D. Wilson$^1$, B. E. Warren$^{1,2}$, J. Irwin$^3$,
      J. H. Knapen$^{4,5}$, F. P. Israel$^6$,
\newauthor   
S. Serjeant$^7$, 
D. Attewell$^1$, 
G. J. Bendo$^8$, 
E. Brinks$^9$, 
H. M. Butner$^{10}$, D. L. Clements$^8$, \newauthor
J. Leech$^{11}$, 
H. E. Matthews$^{12}$, 
    S. M\"uhle$^{13}$, 
A. M. J. Mortier$^{14}$, T. J. Parkin$^1$,  \newauthor
G. Petitpas$^{15}$,
      B. K. Tan$^{11}$, 
R. P. J. Tilanus$^{16,17}$, 
    A. Usero$^{18}$, M. Vaccari$^{19}$, \newauthor
P. van der Werf$^{6}$, T. Wiegert$^{20}$, M. Zhu$^{21}$
    \\
    $^1$Department of Physics \& Astronomy, McMaster University, Hamilton, 
        Ontario L8S 4M1, Canada\\
$^2$ Currently at the International Centre for Radio Astronomy Research,
University of Western Australia \\
    $^3$Department of Physics, Engineering Physics and Astronomy,
        Queen's University, Kingston, Ontario K7L 3N6, Canada\\
    $^4$Instituto de Astrof\'isica de Canarias, E-38200 La Laguna,
    Tenerife, Spain\\
    $^{5}$Departamento de Astrof\'\i sica, Universidad de La Laguna, E-38205 La
Laguna, Tenerife, Spain\\
    $^6$Sterrewacht Leiden, Leiden University, PO Box 9513, 2300 RA Leiden,
        The Netherlands\\
    $^7$Department of Physics \& Astronomy, The Open University, 
        Milton Keynes MK7 6AA, United Kingdom\\
    $^{8}$Astrophysics Group, Imperial College, Blackett Laboratory,
        Prince Consort Road, London SW7 2AZ, United Kingdom\\
    $^{9}$Centre for Astrophysics Research, University of Hertfordshire, 
        College Lane, Hatfield AL10 9AB, United Kingdom\\
    $^{10}$Department of Physics and Astronomy, James Madison
University, MSC 4502 - 901 Carrier Drive, Harrisonburg, VA 22807, U.S.A. \\ 
    $^{11}$Department of Physics, University of Oxford, Keble Road, Oxford OX1
        3RH, United Kingdom\\
    $^{12}$National Research Council Canada, Herzberg Institute of Astrophysics,
        DRAO, P.O. Box 248, White Lake Road, Penticton, \\British Columbia
        V2A 69J, Canada\\
    $^{13}$Joint Institute for VLBI in Europe, Postbus 2, 7990 AA Dwingeloo, 
        The Netherlands\\
$^{14}$Scottish Universities Physics
    Alliance,  Institute for
Astronomy, University of Edinburgh, Royal Observatory, Blackford Hill,\\
Edinburgh, EH9 3HJ, UK\\
    $^{15}$Harvard-Smithsonian Center for Astrophysics, 60 Garden St., MS-78,
        Cambridge, MA 02138, USA\\
    $^{16}$Joint Astronomy Centre, 660 N. A'ohoku Pl., University Park, Hilo, HI 
        96720, USA\\
    $^{17}$Netherlands Organisation for Scientific Research, Laan van Nieuw 
        Oost-Indie 300, NL-2509 AC The Hague, The Netherlands\\
    $^{18}$Observatorio de Madrid, OAN, Alfonso XII, 3, E-28014 Madrid, Spain\\
    $^{19}$Dipartimento di Astronomia, Universit\'a di Padova, Vicolo 
        dell'Osservatorio 5, 35122 Padua, Italy\\
    $^{20}$Department of Physics and Astronomy, University of Manitoba, Winnipeg, 
        Manitoba R3T 2N2, Canada\\
$^{21}$National Astronomical Observatories, Chinese Academy of Sciences,
20A Datun Road, Chaoyang District, Beijing, China\\
}
\begin{document}

\date{11 June 2010}

\pagerange{\pageref{firstpage}--\pageref{lastpage}} \pubyear{2010}

\maketitle

\label{firstpage}

\begin{abstract}
An analysis of large-area CO $J$=3-2 maps from the James Clerk Maxwell
Telescope for 12 nearby spiral galaxies reveals low velocity
dispersions in the molecular component of the interstellar medium.
The three lowest luminosity galaxies show a relatively flat velocity
dispersion as a 
function of radius while the remaining nine galaxies show a
 central peak with a radial fall-off within $0.2-0.4
r_{25}$. 
Correcting for the average contribution due to the internal velocity dispersions
of a population of giant molecular clouds, the average cloud-cloud
velocity dispersion across the galactic disks is $6.1 \pm 1.0$ km
s$^{-1}$ (standard deviation 2.9 km s$^{-1}$), in 
reasonable agreement with previous measurements for the Galaxy and M33.
The cloud-cloud velocity dispersion derived from the CO data is on
average two times smaller than the HI 
velocity dispersion measured in the same galaxies. The low cloud-cloud velocity
dispersion implies that the molecular gas is the critical component
determining the stability of the galactic disk against gravitational
collapse, especially in those regions of the disk which are H$_2$ dominated.
The cloud-cloud velocity dispersion shows a significant positive correlation
with both the far-infrared luminosity, which traces the star
formation activity, and the K-band absolute magnitude, which traces
the total stellar mass.
For three galaxies in the Virgo cluster, 
smoothing the data to a resolution of 4.5 kpc (to match the typical
resolution of high redshift CO observations) increases the measured
velocity dispersion by roughly a factor of two, 
comparable to the
dispersion measured recently in a normal galaxy at $z=1$.
This comparison suggests that the mass and star formation rate
surface densities may be similar in galaxies from $z=0-1$ and that the
high star formation rates seen at
$z=1$ may be partly due to the presence of physically larger molecular
gas disks.
\end{abstract}


\section{Introduction}

The vertical structure of the interstellar medium is determined by a delicate
balance between gravity and pressure. The concentration of
mass in the stellar disk drags the gas into a thin disk whose vertical
scale height is controlled by the velocity dispersion.
The velocity dispersion of the gas is also an important parameter for star
formation laws based on the Toomre Q criterion \citep{t64,k89}.
The atomic phase of the interstellar medium (ISM) 
has been well studied and interpreted, most recently by
\citet[][see below]{t09}.
However, the velocity dispersion of the star forming
molecular gas is much less well understood, although the available
data are consistent with a significantly thinner and dynamically
colder molecular disk \citep{sb89,ws90,cb97}.

Determining the  velocity dispersion of the molecular gas in the
Galaxy is complicated by our location within the plane of the disk and
the difficulty in removing the effects of streaming motions associated
with spiral arms. 
Nearby, relatively face-on galaxies are better targets, provided
sufficient sensitivity and spectral and spatial resolution can be
achieved. However, an additional complicating factor in any analysis is 
the fact that the observed velocity dispersions are not significantly
larger than the internal velocity dispersion of an individual
giant molecular cloud (GMC). For example, \citet{s87} measure internal velocity
dispersions ranging from 1 to 8 km s$^{-1}$  and a relation between
internal velocity dispersion and cloud mass that goes as $M = 2000
\sigma_v^4$. Certainly the net effect of the internal velocity
dispersion of a population of giant molecular clouds in the beam needs
to be considered in an analysis of the cloud-cloud velocity dispersion
of nearby, near face-on galaxies.

For our own Galaxy,
\citet{c85} obtained a one-dimensional cloud-cloud
velocity dispersion of 3.0 km s$^{-1}$, significantly smaller than the
value of 7-9 km s$^{-1}$ obtained by \citet{s84}. 
\citet{sb89} argue that the velocity dispersion measured by
\citet{c85} is actually the internal velocity dispersion of
individual clouds and obtain a cloud-cloud velocity dispersion of $7.8
\pm 0.6$ km s$^{-1}$ for clouds within 3 kpc of the Sun. Although this
measurement includes small-scale streaming motions, they argue that
the true value is only 20\% smaller when the streaming motions are
removed \citep{sb89}.
More recently \citet{sl05,sl06} have re-derived the scale height of
the molecular gas to be 35 pc for clouds less massive than $2\times
10^5$ M$_\odot$ and only 20 pc for clouds more massive than this
limit. This scale height of 20 pc implies a one-dimensional 
cloud-cloud velocity
dispersion of just 4 km s$^{-1}$ for the more massive molecular clouds.
\citet{cb97} point out that, given the observed scale heights of HI
and CO in the Galaxy and a velocity dispersion of 9 km s$^{-1}$ in the
atomic gas, we would expect a cloud-cloud velocity dispersion of only 2.4 km
s$^{-1}$ in the molecular gas, a value which is significantly smaller
than any of the measurements. 

There are relatively few measurements of the velocity dispersion of
the molecular gas in other galaxies. \citet{ws90} used a combination
of interferometric observations of individual giant molecular clouds
with single dish observations of M33 to measure a cloud-cloud velocity
dispersion of $5\pm 1$ km s$^{-1}$. 
\citet{cb97} observed NGC 628 and NGC 3938 using the CO $J$=1-0 and
$J$=2-1 lines and observed velocity dispersions of 6 km s$^{-1}$ and
8.5 km s$^{-1}$, respectively. They used gaussian fits to the observed
line widths and corrected for saturation effects \citep{gb93}.
\citet{w02} made CO $J$=1-0 and $J$=3-2 observations of NGC 6946 and
measured velocity dispersions from second moment maps 
of $8.9 \pm 2.2$ km s$^{-1}$ and $6.0
\pm 1.6$ km s$^{-1}$ in the two lines. Neither of these two 
studies \citep{cb97,w02}  attempted to correct for the internal
velocity dispersion of individual GMCs, and so these measurements are
upper limits to the cloud-cloud velocity dispersion.

There have been a number of theoretical attempts to model the
cloud-cloud velocity dispersion in the Galaxy.
\citet{jo88} propose that cloud-cloud gravitational scattering in a
differentially rotating galactic disk acts to increase the random
kinetic energy of the cloud population. In this model, inelastic collisions
between clouds act as an energy sink, resulting in an equilibrium value
for the 
one-dimensional velocity dispersion of 5-7 km s$^{-1}$.
\citet{g91} extended this work using both analytical and numerical
analyses and obtain a value of 5 km s$^{-1}$ for the two-dimensional
velocity dispersion in the plane of the disk. \citet{t91} performed N
body simulations of clouds and stars which include inelastic cloud
collisions but not close 2-body gravitational encounters. They find
typical velocity dispersions of 3 km s$^{-1}$ which increase in
galaxies with stronger spiral structure. More recently, \citet{tt09}
have developed a three dimensional model including ISM cooling to 300
K with a resolution of 8 pc and find typical velocity dispersions of
10 km s$^{-1}$ for a model of the Milky Way.

Although there have been relatively few measurements of the velocity
dispersion in the molecular gas in galaxies, the velocity dispersion
of the atomic gas has been well studied 
\citep[see][and references therein]{t09}.
Most recently,
the THINGS survey \citep{w08} has produced high-resolution HI maps of
34 spiral and irregular galaxies with distances less than 11 Mpc. 
The typical velocity dispersion in the atomic gas at radii between
$r_{25}/2$ and $r_{25}$ is $11\pm 3$ km s$^{-1}$ 
for galaxies with
inclinations less than 60$^o$ \citep{l08}.
\citet{t09} find an HI line width that falls off systematically with
radius which they link to the energy provided by supernovae linked to
recent star formation. They also find a characteristic HI velocity
dispersion of $10 \pm 2$ km s$^{-1}$ at $r_{25}$, which often marks
the extent of significant star formation in the disk.

In this paper, we present measurements of the velocity dispersion for
the molecular component of the interstellar medium using data from 
the JCMT Nearby Galaxies Legacy Survey (NGLS) as well as a follow-up
program on HI-rich spiral galaxies in the Virgo Cluster.
We use new, wide-area observations of the CO $J$=3-2
emission for 12 large
spiral galaxies  with inclinations $<60^o$ and distances $< 17$ Mpc.
We discuss the
observations, data reduction, and analysis methods used in 
\S\ref{sec-obs}. A more detailed comparison with the results of
\citet{cb97} is given in Appendix A.
In \S\ref{sec-veldisp}, we discuss the observed values of the velocity
dispersion of the CO $J$=3-2 transition and the correction factor
necessary due to the non-trivial internal velocity dispersion of
individual giant molecular clouds. The potential need to
correct for a non-isotropic velocity dispersion in galaxies with
inclinations larger than about 30$^o$ is discussed in Appendix B.
In \S\ref{sec-disc},
we compare our results with recent observations of the velocity
dispersion of the atomic gas component from the THINGS survey
\citep{w08}, investigate correlations of the velocity
dispersion of the molecular gas with global galaxy properties such as
mass and star formation rate, discuss the implications for
understanding disk stability in galaxies, and compare our data with
recent observations of spiral galaxies at $z=1$.
We give our conclusions 
in \S\ref{sec-concl}.

\section{Observations and Data Processing}\label{sec-obs}

\begin{figure*}
\includegraphics[width=64mm]{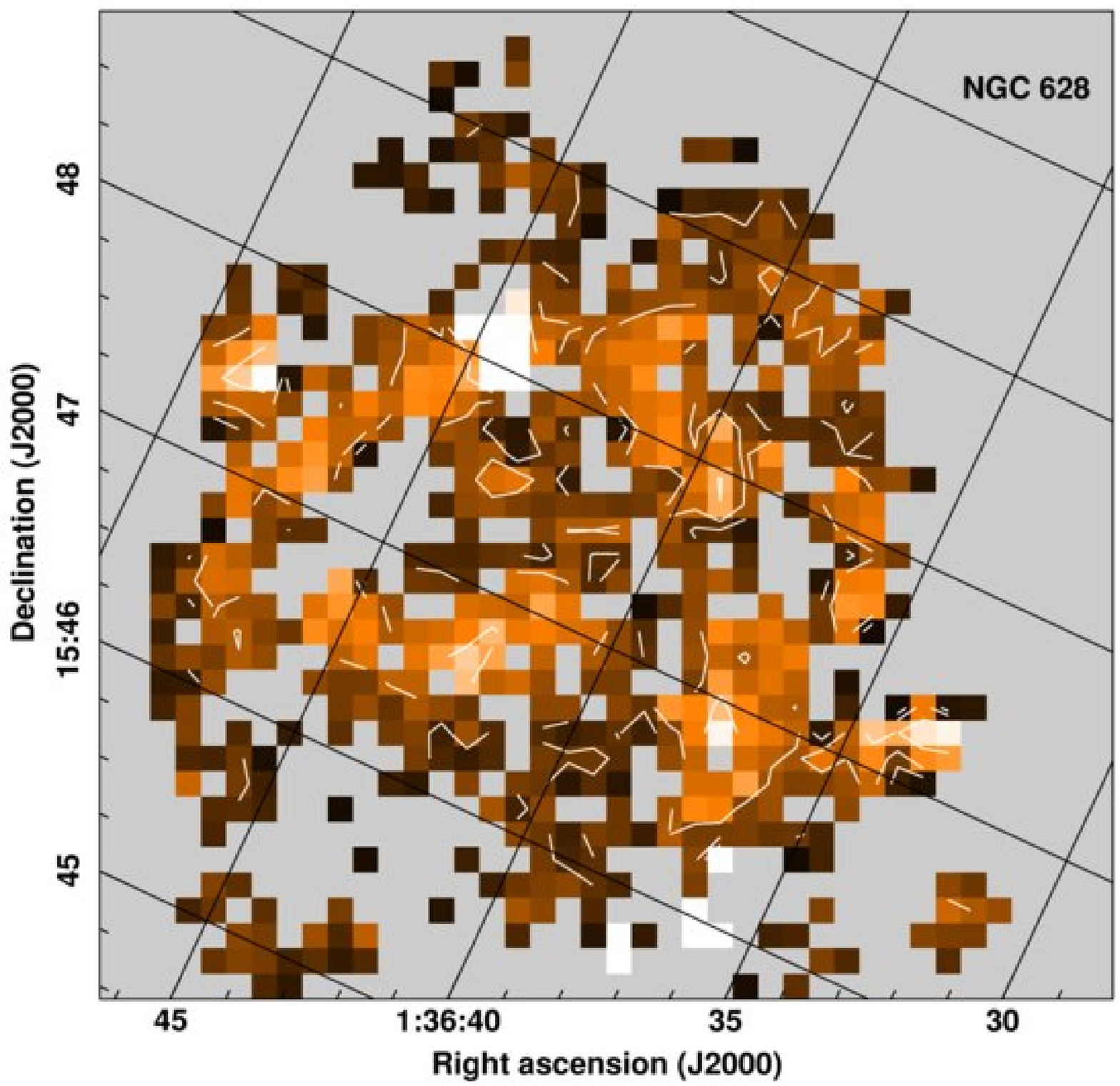}
\includegraphics[width=64mm]{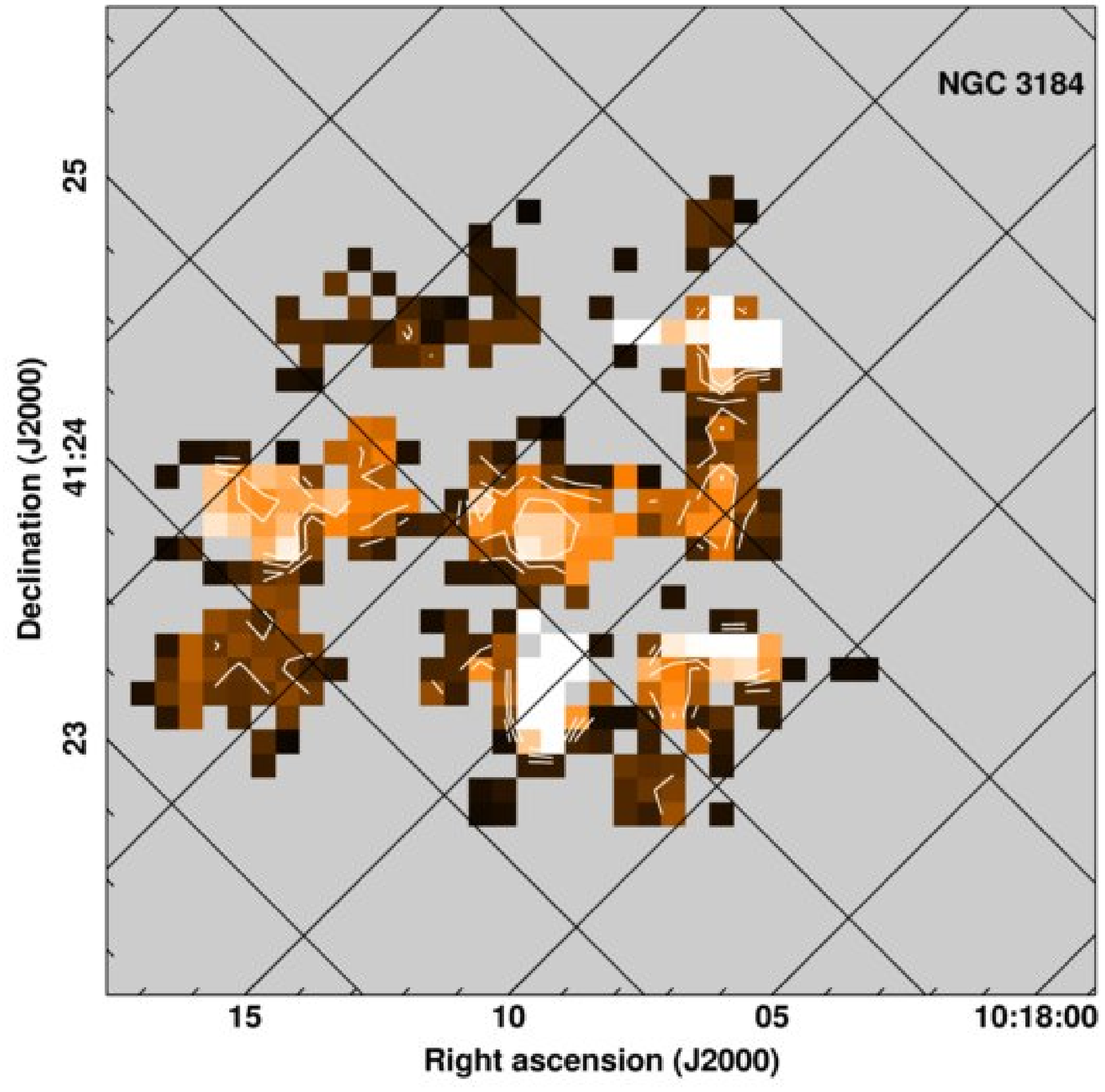}
\includegraphics[width=74mm]{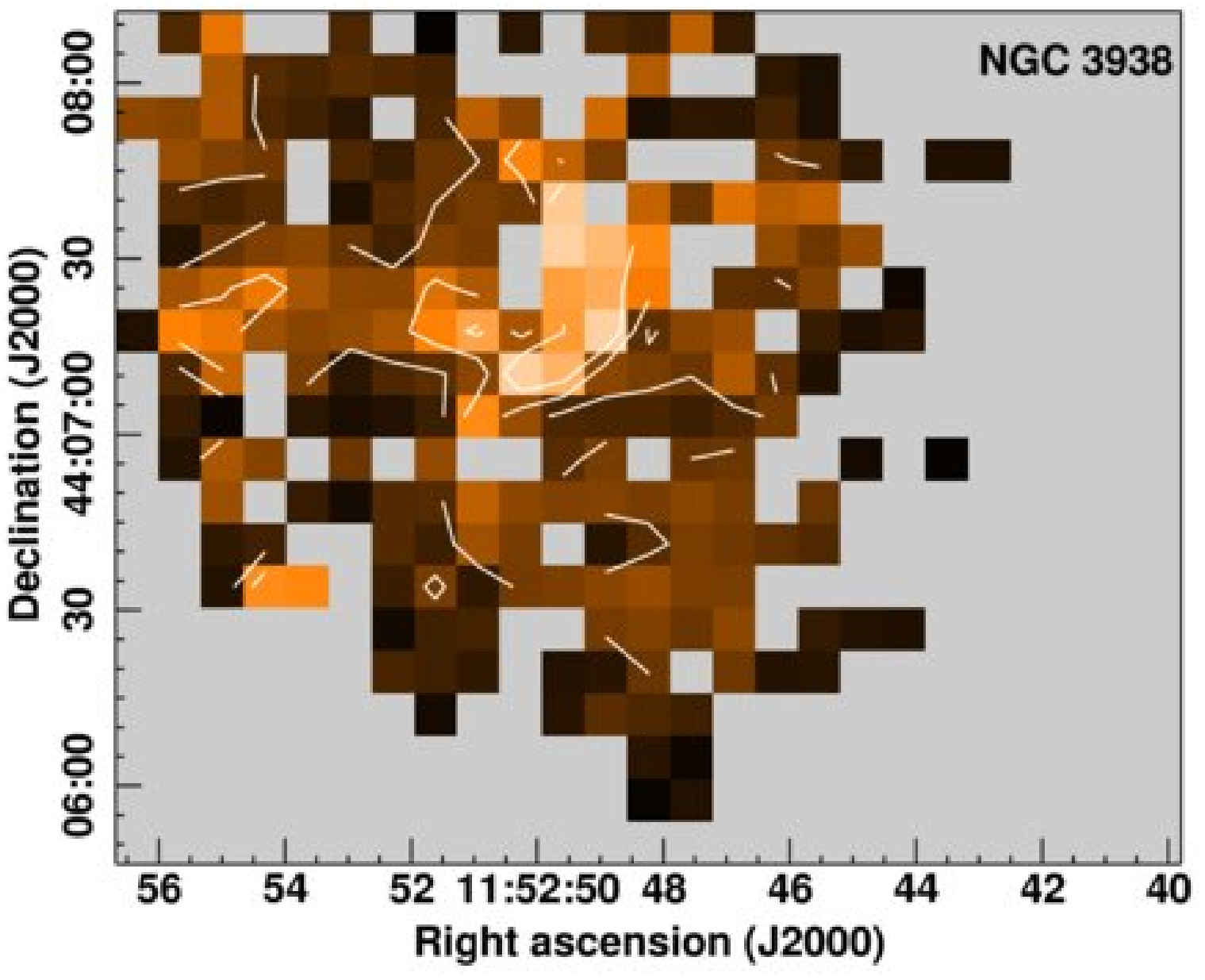}
\includegraphics[width=74mm]{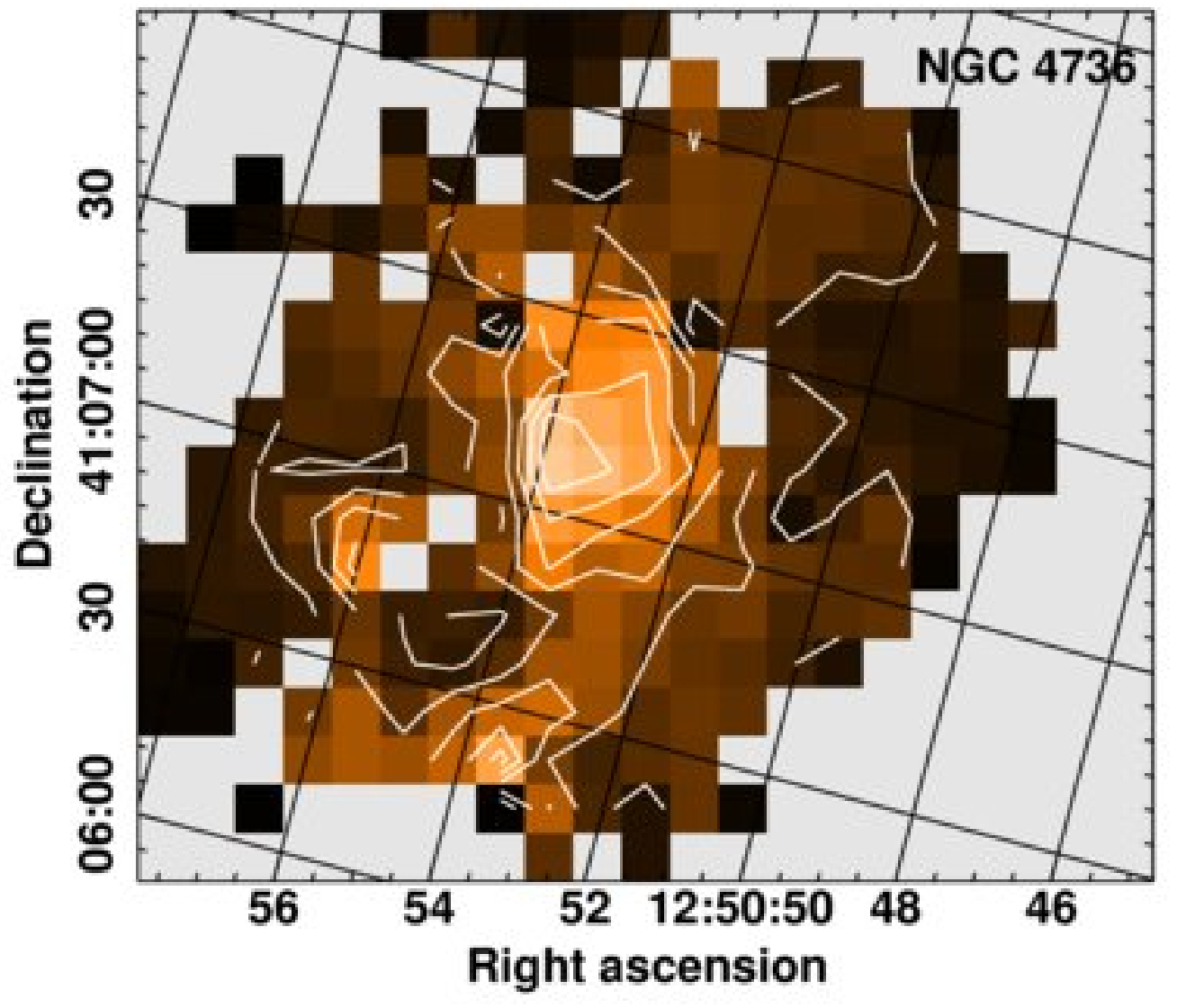}
\includegraphics[width=84mm]{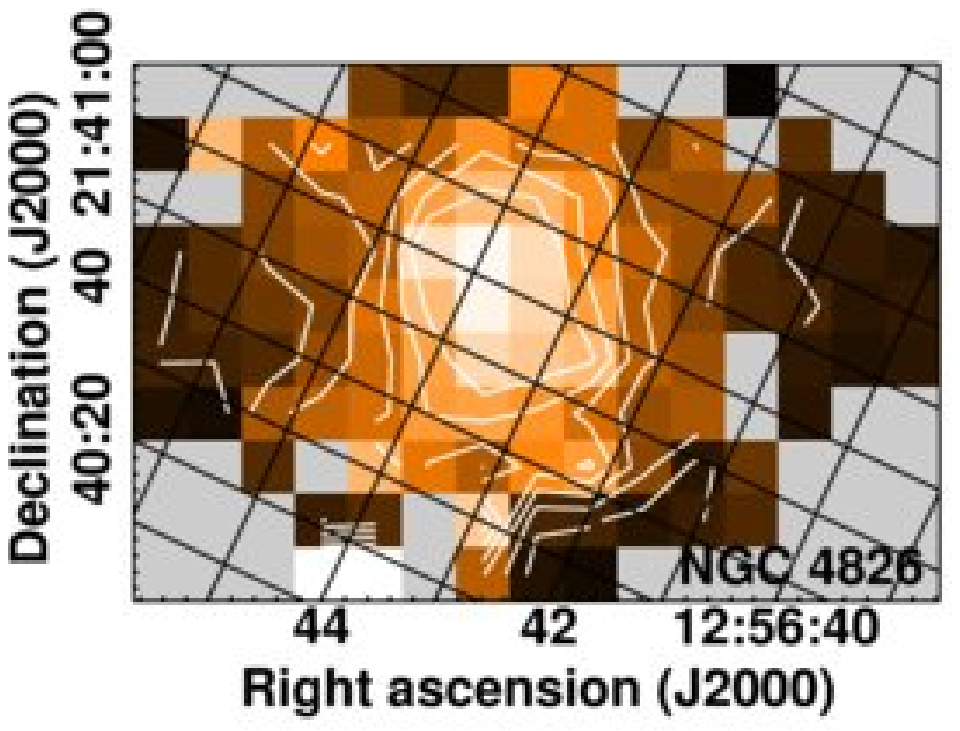}
\includegraphics[width=84mm]{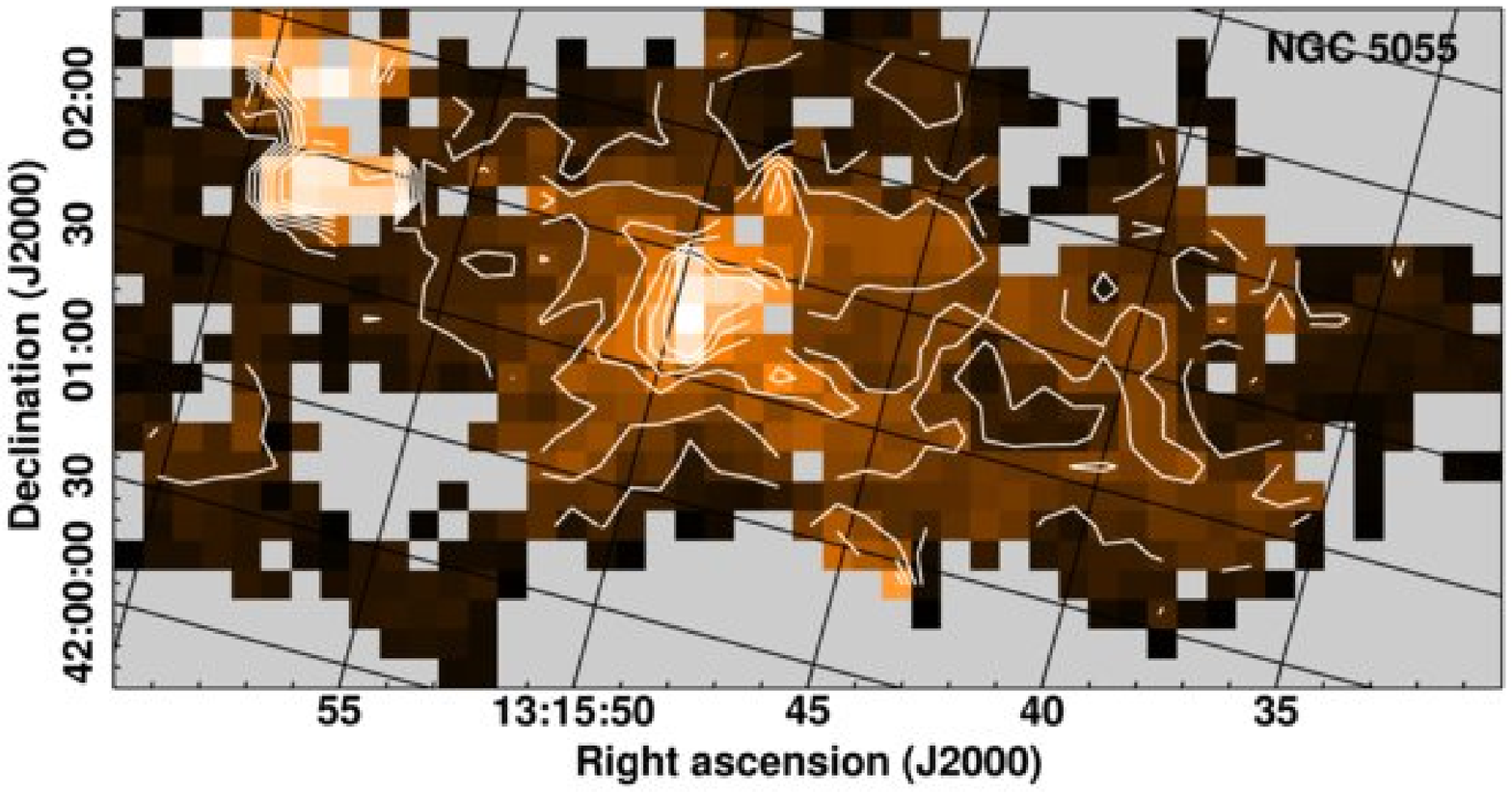}
\caption{CO $J$=3-2 velocity dispersion for six galaxies which are not
  members of the Virgo cluster. 
(a) NGC 628.  Contour levels are 2.5, 5, 10 km
s$^{-1}$ and the colour scale peak is 15 km s$^{-1}$.
(b)  NGC 3184. Contour levels are 5, 10, 15 km
s$^{-1}$ and the colour scale peak is 25 km s$^{-1}$.
(c)  NGC 3938. Contour levels and colour scale are the same as for NGC 3184.
(d)  NGC 4736. Contour levels are 10, 20, ... 60 km
s$^{-1}$ and the colour scale peak is 85 km s$^{-1}$.
(e)  NGC 4826. Contour levels and colour scale are the same as for NGC 4736.
(f) NGC 5055. Contour levels and colour scale are the same as for NGC 4736.
\label{fig-veldisp}}
\end{figure*}

\begin{figure*}
\includegraphics[width=64mm]{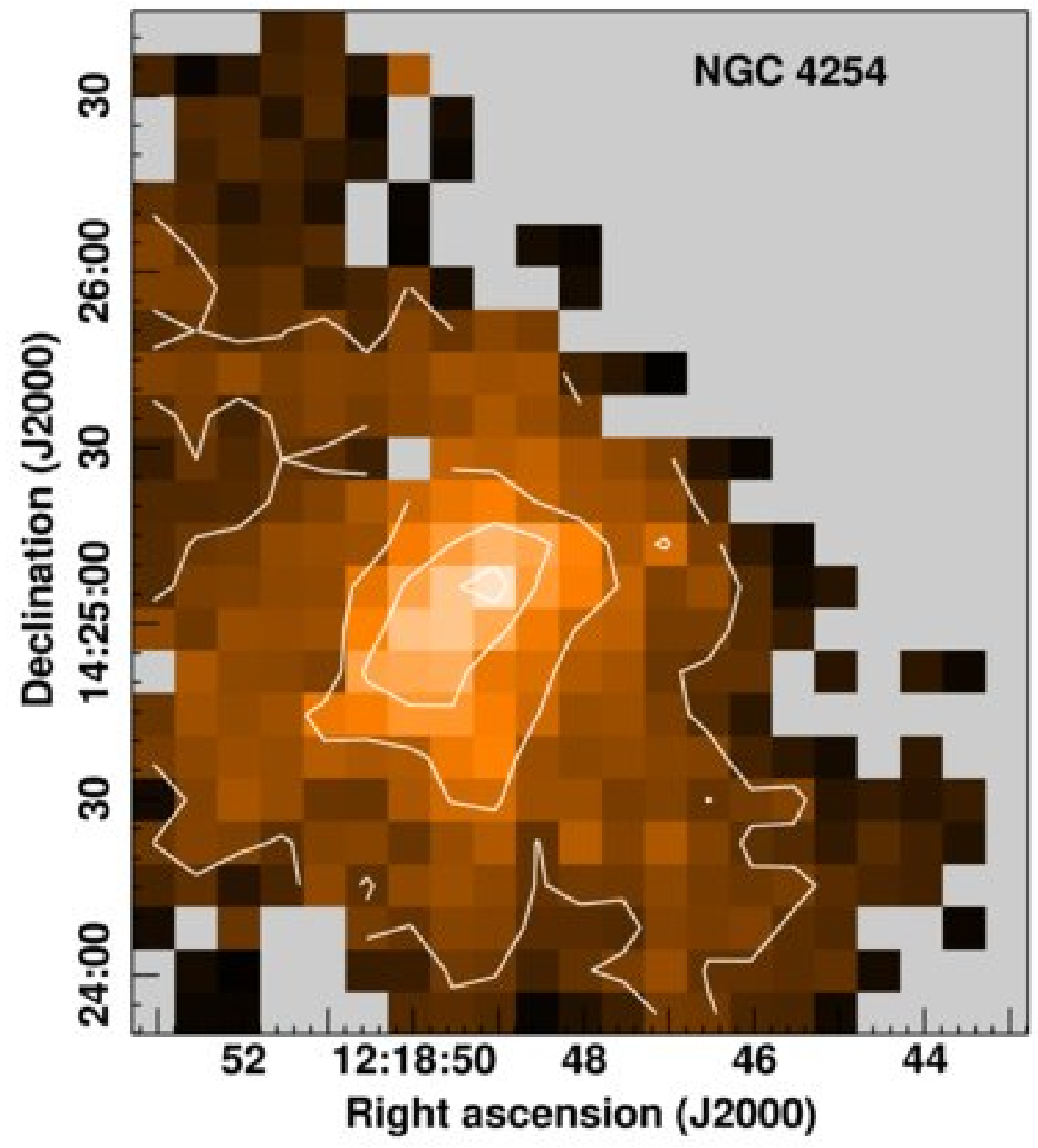}
\includegraphics[width=64mm]{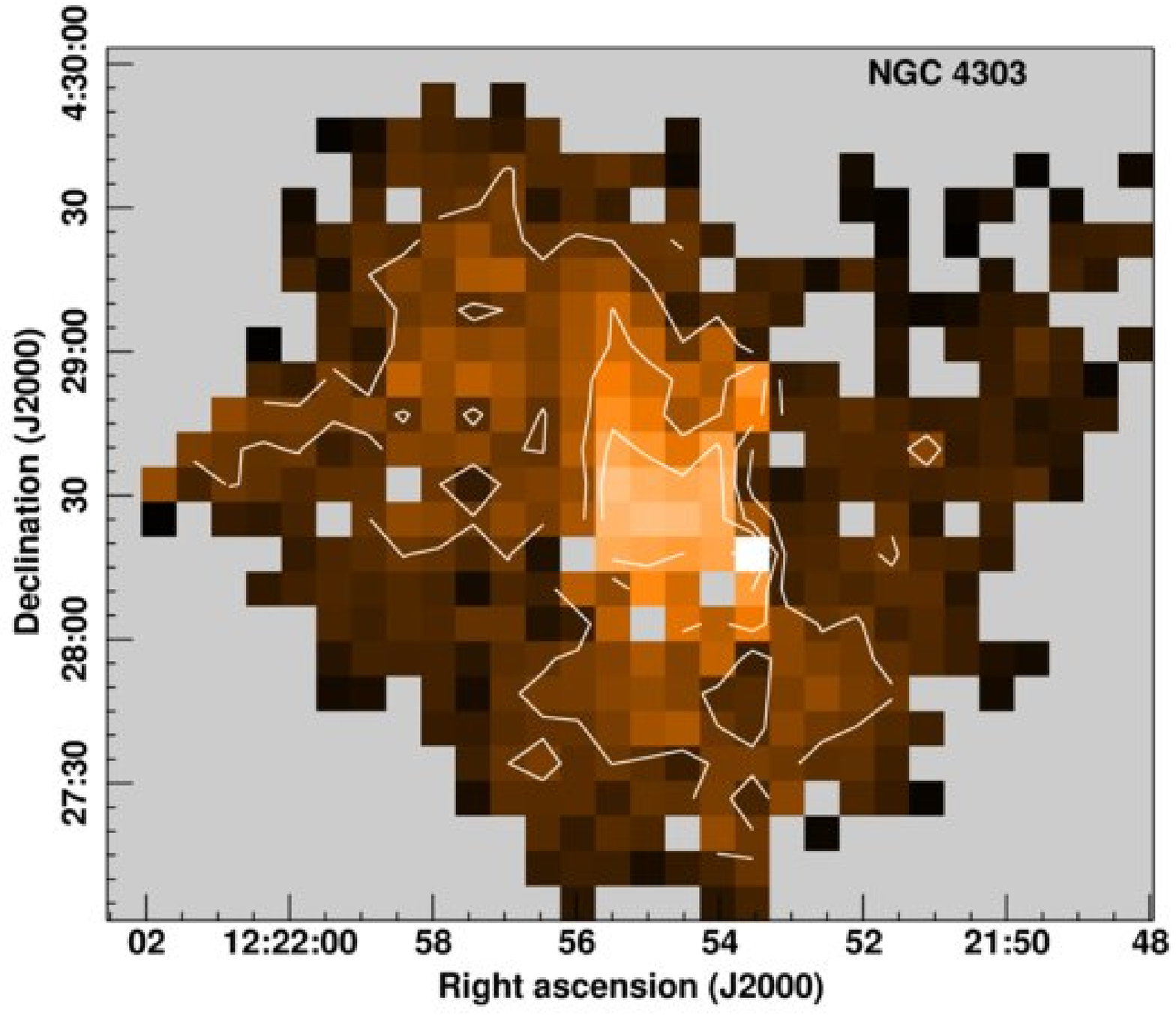}
\includegraphics[width=84mm]{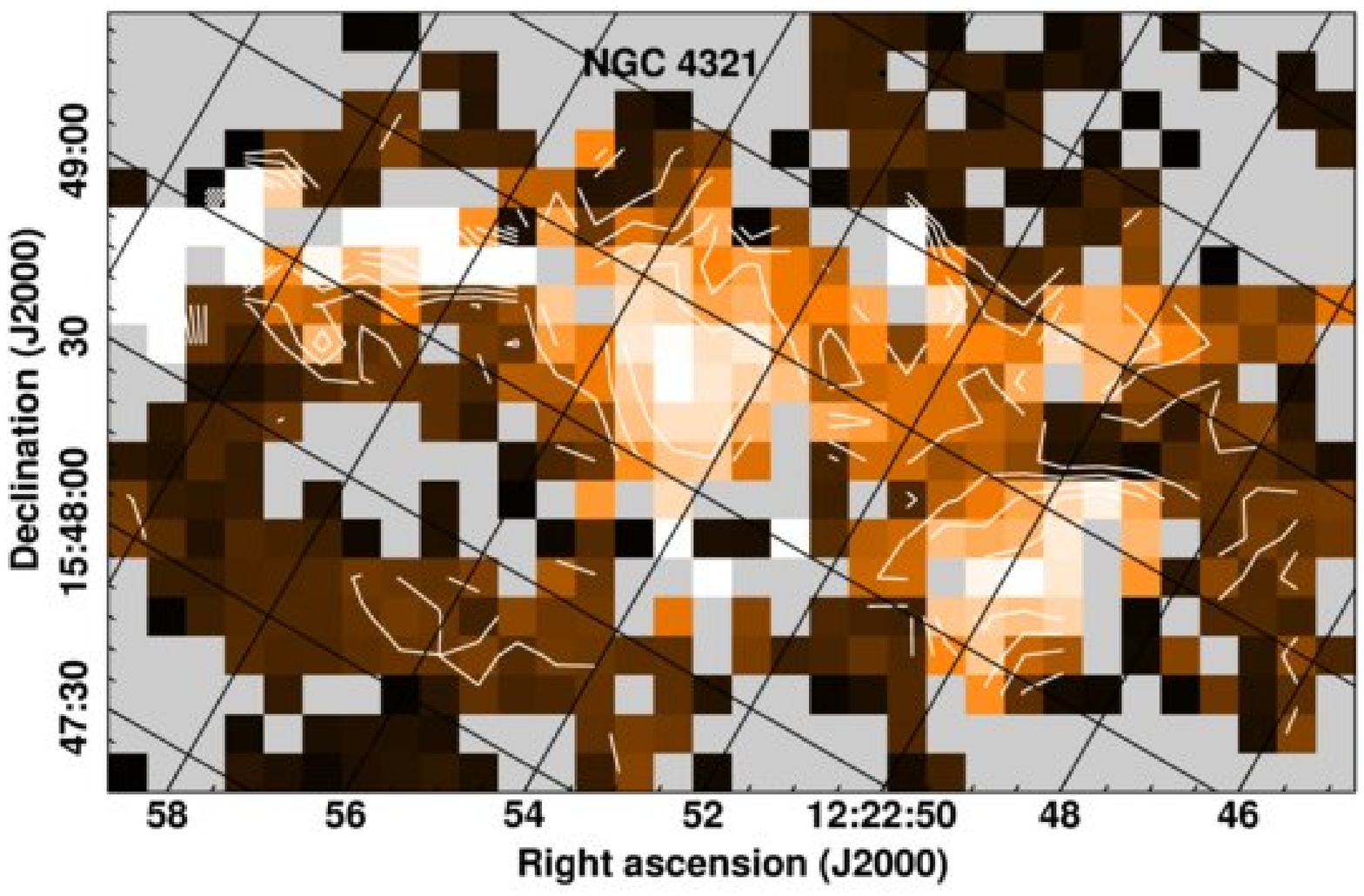}
\includegraphics[width=84mm]{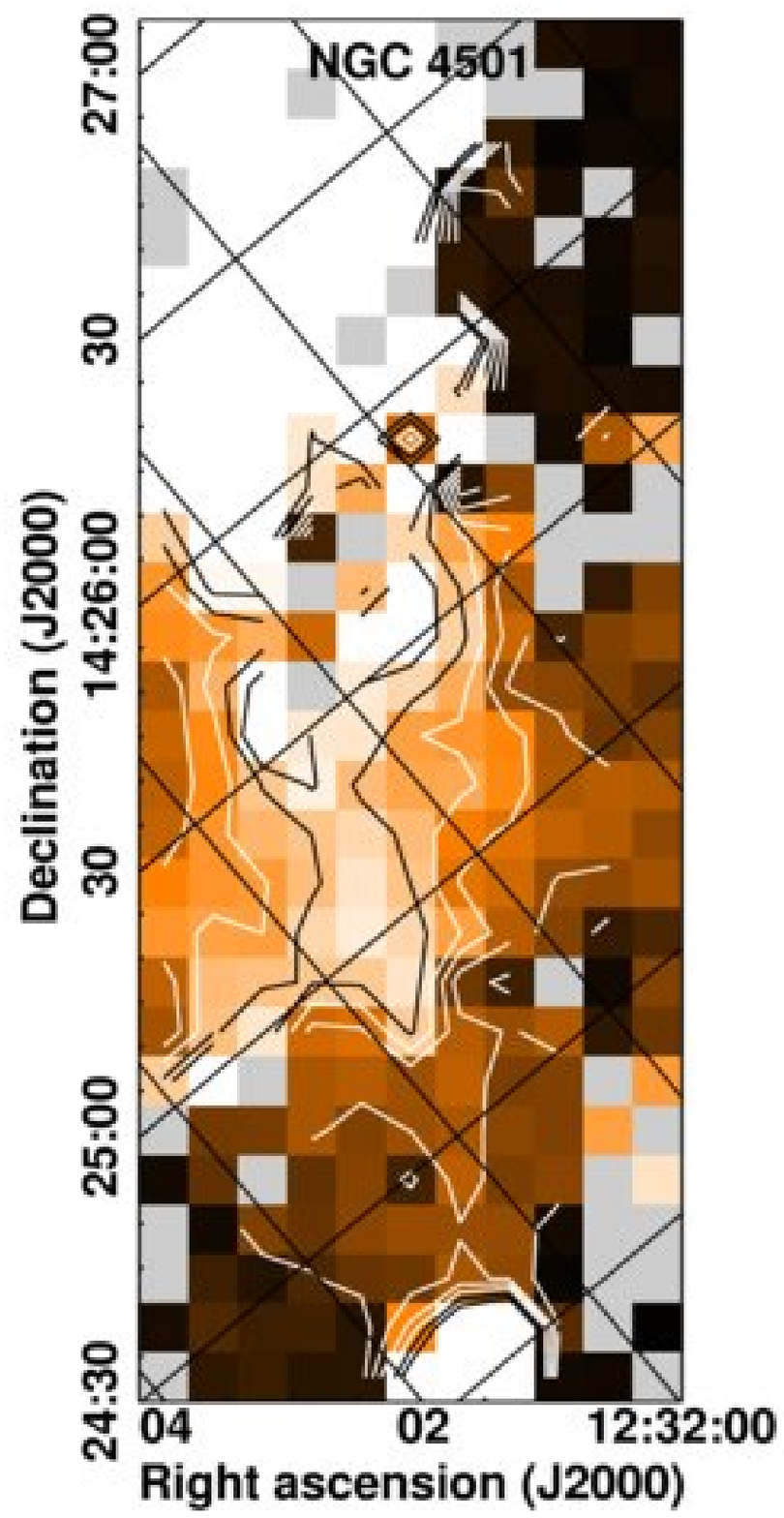}
\includegraphics[width=84mm]{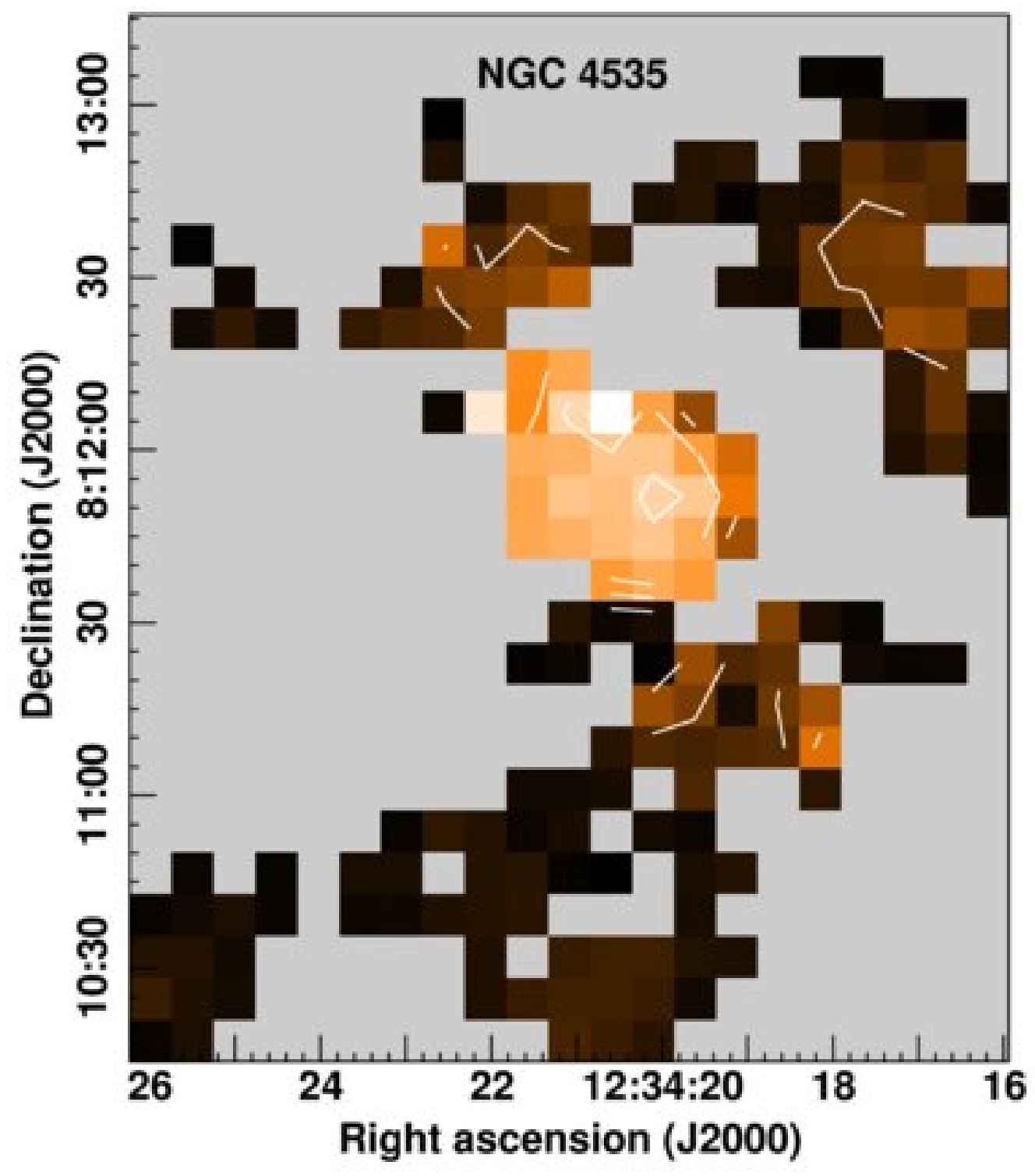}
\includegraphics[width=84mm]{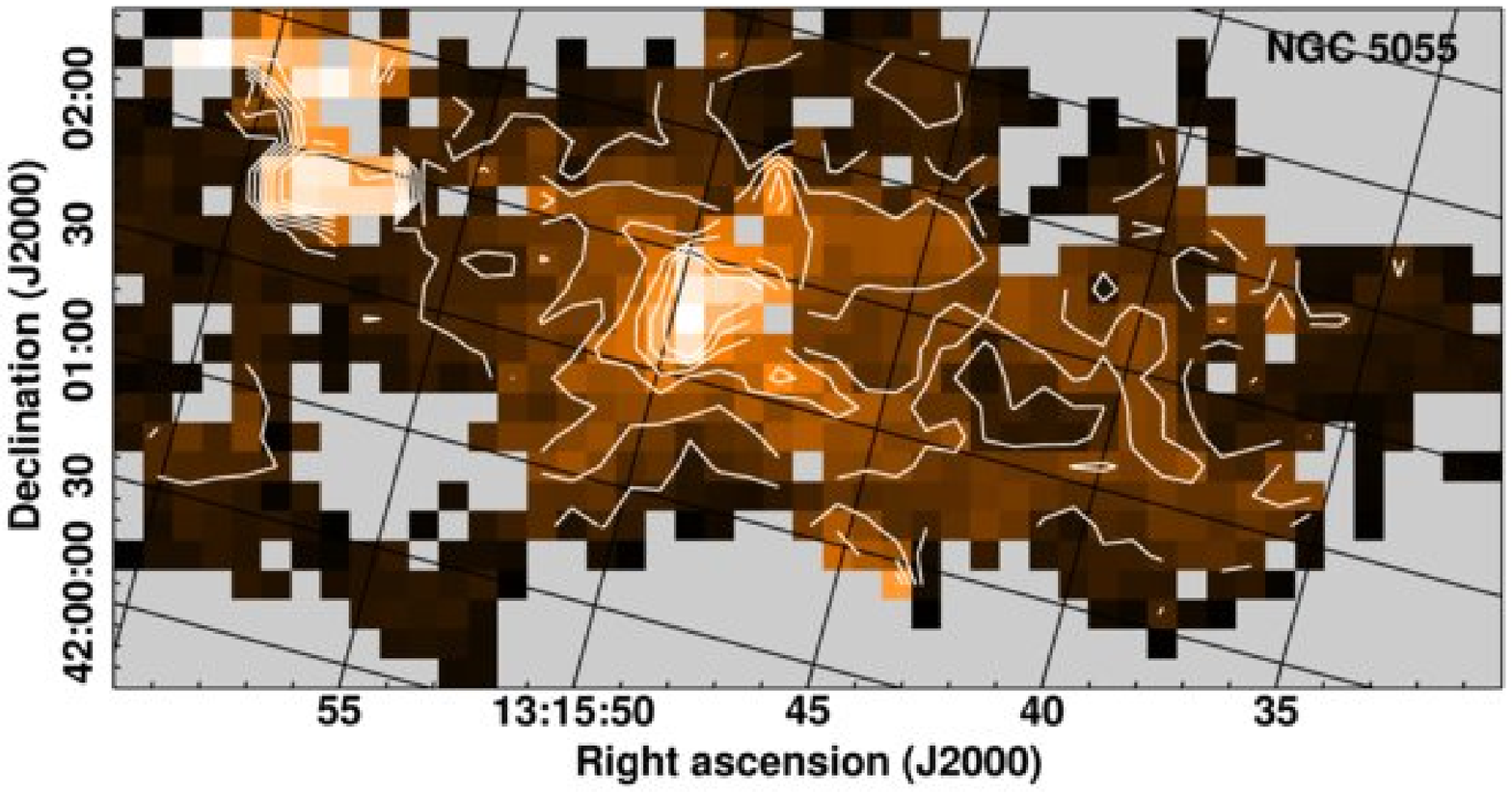}
\caption{CO $J$=3-2 velocity dispersion for five galaxies
  which are  
  members of the Virgo cluster and NGC 2403, which is the closest
  galaxy in our sample. Colour scale
  runs from 0 (black) to 50 km s$^{-1}$ (white)
  and contour levels are 10, 20, 30, 40, 50 km s$^{-1}$  unless
  otherwise noted. 
(a) NGC 4254.
(b)  NGC 4303.
(c)  NGC 4321. 
(d) NGC 4501. Contour levels are 10, 20, ... 60 km
s$^{-1}$ and the colour scale peak is 65 km s$^{-1}$.
(e) NGC 4535. 
(f) NGC 2403. Contour levels are 5, 10, 15 km s$^{-1}$
and the colour scale peak is 25 km s$^{-1}$.
\label{fig-veldisp2}}
\end{figure*}

\subsection{JCMT CO $J$=3-2 data}

\renewcommand{\thefootnote}{\alph{footnote}}
\begin{table*}
 \centering
 \begin{minipage}{140mm}
  \caption{Galaxy Properties and Velocity Dispersions\label{tbl-props}}
  \begin{tabular}{lccccccccccc}
  \hline
Galaxy & $V_{hel}$ & Type\footnotemark[1]
& $D_{25}$\footnotemark[1]
& $i$\footnotemark[1]
& $PA$\footnotemark[2]
& $K(tot)$\footnotemark[3]
& $\log L_{FIR}$\footnotemark[4]
& Distance\footnotemark[5]
& $\sigma_{obs}$\footnotemark[6]
 & $\sigma_{c-c}$\footnotemark[7]
 & $\sigma_{HI}$\footnotemark[8] \\
 & (km s$^{-1}$) 
&  & ($^{\prime}$) & ($^o$) & ($^o$) & (mag) %& (M$_\odot$ yr$^{-1}$) 
& (L$_\odot$) & ($Mpc$)  & (km s$^{-1}$) & (km s$^{-1}$) & (km s$^{-1}$) 
  \\
\\
 \hline
NGC 628 & 648 & SA(s)c & 10.5 & 7 & 20 & 6.84 & 9.55 & 7.3 & $4.1\pm 2.5$
 & $3.5\pm 0.1$   & 12 \\ 
%2.5$ & 3.5  & 11.8 \\ 
NGC 2403 & 121 & SAB(s)cd & 21.9 & 56 & 124 & 6.19 & 9.10 & 3.2 &
$5.2\pm 4.9$ & $3.9\pm 0.2$  & 14 \\ 
%$5.2\pm 4.9$ & 3.9  & 14.0 \\ 
NGC 3184 & 574 & SAB(rs)cd & 7.4 & 16 & 179 & 7.78
& 9.69 & 11.1 & $6.8\pm 5.9$ & $5.8\pm0.4$  & 13 \\
%& 9.69 & 11.1 & $6.8\pm 5.9$ & 5.8  & 13.1 \\
NGC 3938 & 829 & SA(s)c & 5.4 & 25 & 52 & 7.81 &
9.72 & 14.2 & $4.4\pm 2.9$ & $2.7\pm 0.3$  & 10\footnotemark[9] \\
NGC 4254 & 2412 & Sa(s)c & 5.4 & 29 & 56 & 6.93
& 10.62 & 16.7 & $9.2 \pm 4.5$ & $8.5\pm 0.3$  & ... \\
NGC 4303 & 1577 & SAB(rs)bc & 6.5 & 17 & 318 & 6.84
& 10.59 & 16.7 & $8.0 \pm 4.2$ & $7.2\pm 0.2$ & ... \\
NGC 4321 & 1599 & SAB(s)bc & 7.6 & 32 & 143 & 6.59
%&  10.47 & 16.7 & $11.6\pm 10.7$ & 11.1  & ... \\
&  10.47 & 16.7 & $12\pm 11$ & $11.1 \pm 0.6$  & ... \\
NGC 4501 & 2213 & SA(rs)b & 6.9 & 58 & 140 & 6.27 &
%10.41 & 16.7 & $ 20.1 \pm 14.9 $ & 19.8 & ... \\
10.41 & 16.7 & $ 20 \pm 15 $ & $20\pm 2$ & ... \\
NGC 4535 & 1972 & SAB(s)c & 7.1 & 41 & 0 & 7.38 & 10.12 & 16.7 & $ 5.1
\pm 4.4 $ & $3.7\pm 0.4$ & ... \\ 
NGC 4736 & 295 & (R)SA(r)ab & 11.2 & 36 & 296 &
%5.11 & 9.79 & 5.2 & $12.4 \pm 8.6 $ & 11.9 & 25.0 \\
5.11 & 9.79 & 5.2 & $12 \pm 9 $ & $11.9 \pm 0.7$ & 25 \\
NGC 4826 & 390 & (R)SA(rs)ab & 10.0 & 57 & 121 &
%5.33 & 9.85 & 7.5 & $ 13.5 \pm 11.2 $ & 13.0 & 60.0 \\
5.33 & 9.85 & 7.5 & $ 14 \pm 11 $ & $13 \pm 2$ & 60 \\
NGC 5055 & 495 & SA(rs)bc & 12.6 & 59 & 102 & 5.61 &
%10.30 & 8.0 & $9.4\pm 6.6$ & 8.7 & 23.9 \\
10.30 & 8.0 & $9.4\pm 6.6$ & $8.7\pm 0.3$ & 24 \\
\hline
\end{tabular}
\begin{tabular}{l}
\footnotemark[1] \citet{buta07} \\
\footnotemark[2] Position angle used for radial averages. NGC 628, NGC
3184: \citet{tam08}; NGC 2403, NGC 4736, NGC 4826, \\ NGC 5055:
\citet{dB09};
NGC 3938: \citet{p00}; NGC 4254, NGC 4303, NGC 4501, NGC 4535: \\
\citep{c90};
NGC 4321: \citet{k93}. \\
\footnotemark[3] K-band total apparent magnitude from \citet{j03}. \\
\footnotemark[4] From \citet{s03} adjusted for the distances adopted here.\\
\footnotemark[5] NGC 628: \citet{k04}; NGC 2403: \citet{f01};
NGC3184: \citet{l02};
NGC 3938, \\ NGC 5055: distance calculated using Hubble flow distance with velocity
corrected for Virgo infall \citep{m00} and \\ $H_o = 73$ km s$^{-1}$ Mpc$^{-1}$;
Virgo Cluster: \citet{m07}; 
NGC 4736, NGC
  4826: \citet{t01}. \\
\footnotemark[6] Observed CO $J$=3-2 velocity
  dispersion and standard deviation. Regions of high velocity
  dispersion in the outer portions \\ of the disk were masked before
  calculating the velocity dispersion (see text).
Excludes a 5.3 kpc diameter
  region in the centre of each \\ Virgo galaxy (9 pixels), NGC 3938
  (11 pixels) and NGC 5055
(19 pixels). All pixels are included for NGC 628, NGC2403, 
\\ and NGC 3184. For NGC 4736 and NGC 4826, a 9 pixel region is excluded
to give an upper limit to the outer disk velocity dispersion \\ (see
text). 
Note that the uncertainty in the mean for
$\sigma_{obs}$ is the same as for $\sigma_{c-c}$.\\
\footnotemark[7] Mean cloud-cloud velocity dispersion and the uncertainty in
the mean corrected for the
    contribution from the internal \\ velocity dispersion of individual
    giant molecular clouds (see text).  Note that the uncertainty in
    the mean is calculated from the \\ standard deviation given in the
    previous column by dividing by
$\sqrt{N}$, where $N$ is the number of
measurements for a given galaxy. \\
\footnotemark[8] HI velocity dispersion measured from naturally
weighted maps from 
  \citet{w08} over same area as the CO \\ measurement. \\
\footnotemark[9]From \citet{sk82}.
\end{tabular}
\end{minipage}
\end{table*}

The CO $J$=3-2 observations were obtained as part of the JCMT Nearby
Galaxies Legacy Survey
(NGLS)\footnote{http://www.jach.hawaii.edu/JCMT/surveys/}, which is
observing an HI flux limited sample of 155 galaxies within 25 Mpc
\citep{w09}.
The angular resolution of the JCMT at this frequency is
14.5$^{\prime\prime}$, which corresponds to a linear resolution
ranging from 0.2 to 1.2 kpc for the galaxies in our sample.
From the large ($D_{25} > 4^\prime$) spiral galaxies observed by the NGLS, we
selected 9 galaxies with inclinations smaller than 60$^o$ to study the
velocity dispersion. Eight additional large galaxies from the NGLS with low
inclinations (IC 2574, UGC 04305, NGC 3031, NGC 3351, NGC 4450, NGC
4579, and NGC 
4725) did not have sufficiently strong or extended detections in 
CO J=3-2 to be useful for this analysis. To this sample we added
three spiral galaxies in the Virgo Cluster (NGC 4303, NGC 4501, and
NGC 4535) which are not part of the NGLS
survey but which have been observed in a similar manner. (A fourth
Virgo Cluster galaxy, NGC 4548, did not have a strong enough detection
to be useful here.) These galaxies were observed as part of a
follow-up program to the NGLS (JCMT proposal M09AC05, PI C. Wilson) to
obtain CO J=3-2 observations to complete the sample of Virgo spiral
galaxies with HI fluxes $> 6.3$ Jy km s$^{-1}$. The data for these
three galaxies were obtained
between 2009 February 14 and 2009 May 26.

Relevant properties of the 12 galaxies are given in Table~\ref{tbl-props}.
All 12 galaxies were
observed in raster mapping mode to cover a rectangular area
corresponding to $D_{25}/2$ with a 1 sigma sensitivity of better than
19 mK ($\rm T_A^*$) at a spectral resolution of 20 km
s$^{-1}$.  We used the
16 pixel array receiver 
HARP-B \citep{bu09} with the ACSIS correlator configured to have a
bandwidth of 1 GHz and a resolution of 0.488 MHz (0.43 km s$^{-1}$
at the frequency of the CO $J$=3-2 transition). 
The combination of uniquely high velocity resolution and
large mapping area is critical for accurate measurements of the
velocity dispersion in the molecular interstellar medium.

Details of the reduction of the CO $J$=3-2 data are given in
\citet{w09} and \citet{w10} and so we discuss here only the
most important 
processing steps and those 
steps that differed from the previous analysis.
The individual raw data files were flagged to remove data from
any of the 16 individual receptors  with bad
baselines and then the scans were combined into a data cube using a
${\rm sinc}(\pi x){\rm sinc}(k\pi x)$ kernel as the weighting function
to determine the contribution of individual receptors to each pixel in
the final map. The pixel size in the maps is 7.276$^{\prime\prime}$.
A mask was created to identify
line-free regions of the data cube and a first-order baseline was fit
to those line-free regions and subtracted from the cube.

We then used
the clumpfind
algorithm \citep{w94} implemented as part of the
CUPID\footnote{CUPID %and KAPPA are 
is part of the Starlink \citep{c08}
  software package, which is available for download from
http://starlink.jach.hawaii.edu} \citep{b07} task 
findclumps to identify regions with emission with signal-to-noise
greater than  2.5
in a data cube that had been boxcar smoothed
by 3x3 spatial pixels and 25 velocity channels. 
%These smoothing parameters and
%signal-to-noise cutoff were chosen to provide a reasonable match to
%the smoothing and cutoffs used in the THINGS survey \citep{w08} so as to
%allow a more direct comparison between the CO and HI velocity
%dispersions. 
Moment maps were created from
the original data cube after applying the mask created by findclumps.
The moment maps which are the focus of this paper are the moment 2
maps, which measure the velocity dispersion, $\sigma_v$, for each
pixel in the image using
$$\sigma_v = \sqrt{\Sigma T_i (v_i - \overline{v})^2 / \Sigma T_i }$$
where $v_i$ and $T_i$ are the velocity and temperature of a given
velocity channel  and $\overline{v}$ is the intensity weighted mean
velocity of that pixel. This method of calculating the velocity
dispersion differs from that used by
\citet{cb97}, who  fit gaussian profiles to
the CO lines. However, in the limit of
gaussian lines with a high signal-to-noise ratio, the values from the
moment 2 maps should agree with the results from fitting a gaussian
directly to the line profiles. A more detailed comparison of our data
with \citet{cb97} is given in Appendix A.
Moment 2 maps are often used to
determine the velocity dispersion in the atomic gas, although this
method is most reliable for simple line shapes, as warped
disks and HI at large scale heights can distort the line profiles
\citep{dB09}.

Because we were using a relatively low signal-to-noise threshold in
creating our masks, the final moment 2 maps sometimes appeared to
contain spuriously high values typically in the outer portions of the
map.  These regions appear white in Figures~\ref{fig-veldisp}
and~\ref{fig-veldisp2}. 
These values would bias the dispersions upward if they were
included in our averages. For 5 galaxies in our sample,
 (NGC 4321, NGC
628, NGC 2403, NGC 3184, and NGC 4826), 
we applied an upper
threshold to remove pixels with values that were higher than the
highest value seen in the central part of the galaxy. The thresholding
values used were 15 km s$^{-1}$ for NGC 628, 25 km s$^{-1}$ for NGC
2403 and NGC 3184, 52 km s$^{-1}$ for NGC 4321, and 85 km s$^{-1}$ for
NGC 4826.
We also applied
a threshold cut of 65 km s$^{-1}$ to NGC 4501 but very high values
persisted in the 
north-western portion of the map (see below); the velocity dispersions
in Table~\ref{tbl-props} are measured in the south-east portion of the map only.
For NGC 5055, we
used a targeted threshold to remove the block of high values seen in the
north-east portion of the map. 

\subsection{HI data from THINGS}

For the 6 galaxies (NGC 628, NGC 2403, NGC 3184, NGC 4736, NGC 4826,
and NGC 5055) that are in common between our sample and the
THINGS sample \citep{w08}, we can compare the velocity dispersions in
the atomic and molecular gas directly.
The moment 2 maps for the THINGS sample were produced using a
slightly different masking technique \citep{w08} than the one we
adopted for our CO analysis. In the THINGS
processing, the data are
first smoothed to a resolution of 30$^{\prime\prime}$. A mask is then
made keeping only those pixels and velocity channels (width
either 2.6 or 5.2 km s$^{-1}$) where the
emission exceeds 2$\sigma$ in 3 adjacent velocity channels. 
We used the natural weighted maps of the
HI velocity dispersion and measure the dispersion over the same region
in which the CO dispersions have been measured. These velocity
dispersions are also given in Table~\ref{tbl-props}. Note that these
HI dispersions are somewhat larger than the typical value of 
$11\pm 3$ km s$^{-1}$ given in \citet{l08} because they are measured
in the inner rather than the outer disks of the galaxies.

\section{Velocity dispersions in the molecular gas disk}\label{sec-veldisp} 

\begin{figure*}
\includegraphics[angle=-90,width=168mm]{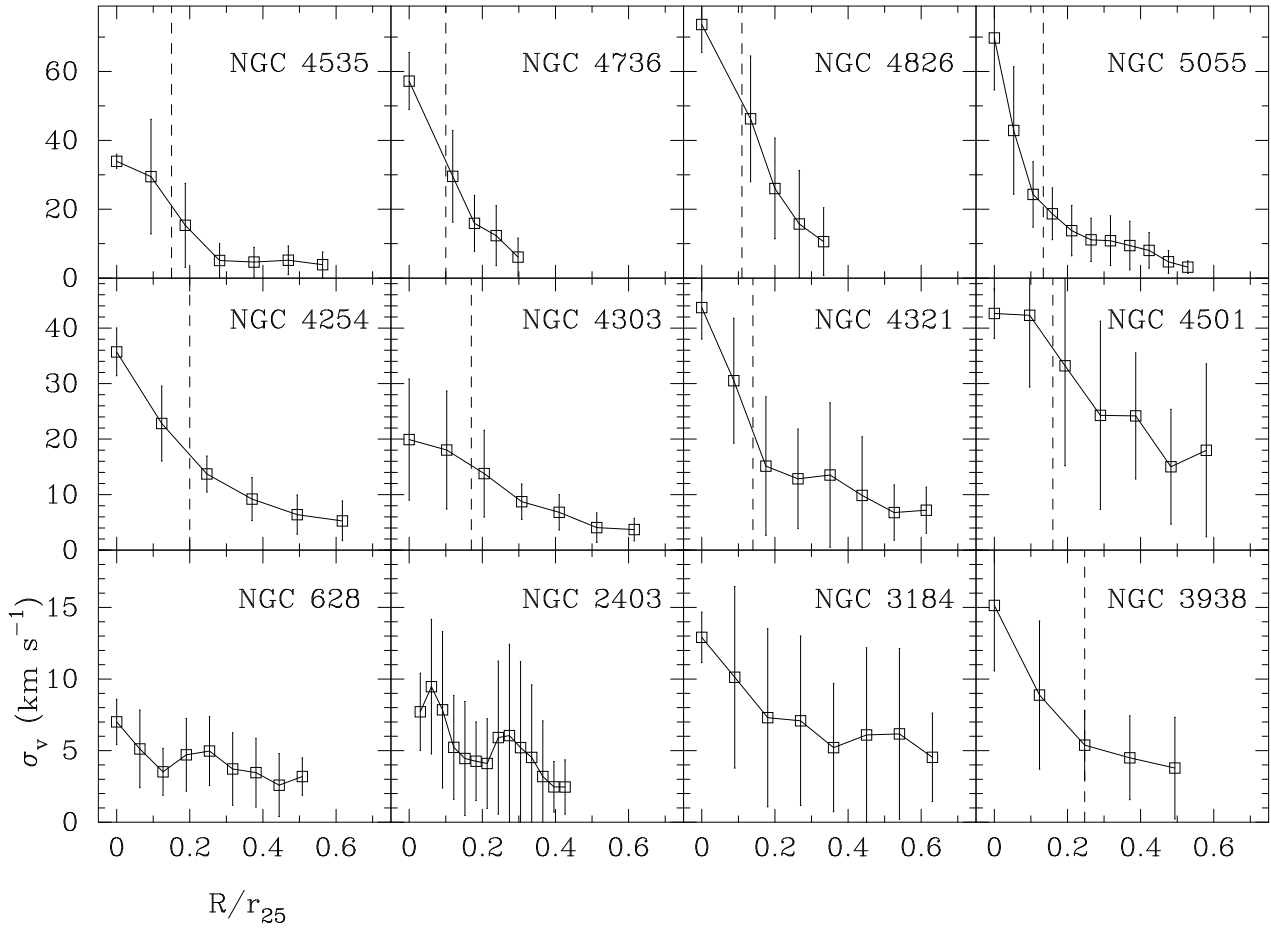}
\caption{Observed CO $J$=3-2 velocity dispersion as a function of
  radius normalized by $r_{25}$. These velocity dispersions have not
  been corrected for any contribution from the internal velocity
  dispersion of individual giant molecular clouds (see text). Where
  appropriate, the central region excluded from calculating the
  disk-averaged velocity dispersion is indicated by the dashed line.
\label{fig-radial}}
\end{figure*}

\begin{figure*}
\includegraphics[angle=-90,width=120mm]{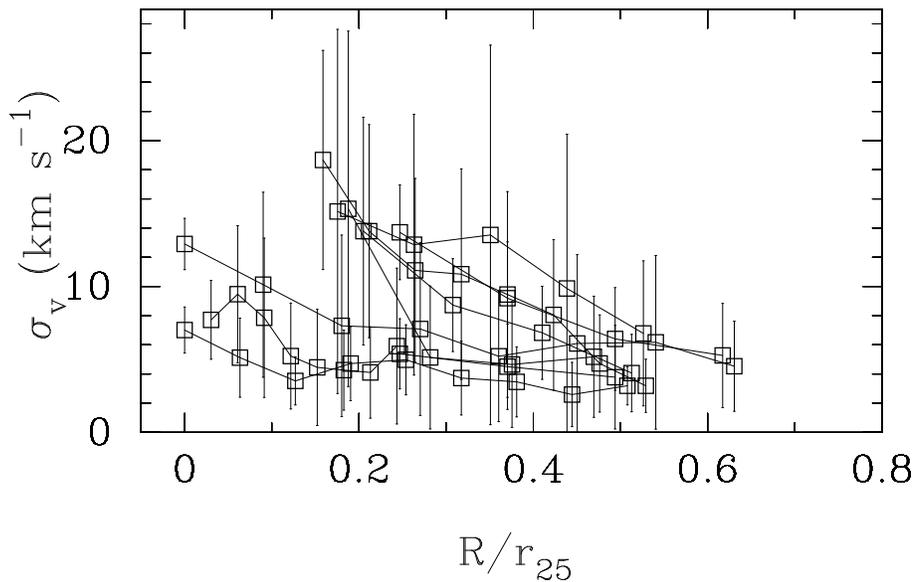}
\caption{Observed CO $J$=3-2 velocity dispersion as a function of
  radius normalized by $r_{25}$, showing nine galaxies on a
  single plot. Data points with $r < 2.65 $ kpc are not plotted for the
  6 galaxies with central peaks in the velocity dispersion.
\label{fig-radplot_all}}
\end{figure*}

The observed CO velocity dispersion maps are shown in
Figures~\ref{fig-veldisp} and~\ref{fig-veldisp2}. 
Figures~\ref{fig-radial} and~\ref{fig-radplot_all}
show the observed velocity dispersion as a function of
radius for the 12 galaxies in our sample. Beam smearing is expected to
have an effect only on the measured 
 velocity dispersions for the central pixel in each plot. 
Three of the galaxies (NGC 628, NGC 2403, and NGC 3184) show
very flat profiles as a function of radius. 
%One galaxy, NGC 2403,
%shows a weak peak in the central velocity dispersion with very limited
%radial extent (possibly produced by beam smearing), while 
These late type spirals are also the least luminous and so presumably
are also the least massive galaxies in our sample.
The remaining 9
galaxies show a central peak in the velocity dispersion extending
 to $0.2-0.4 r_{25}$. 
%Central
% peaks in the velocity dispersion 
%are more likely (but not exclusively) found among the early type than
%the late type spirals. At the two sigma level, the galaxies with
%central peaks in the velocity dispersion tend to have higher
%far-infrared luminosities and be of earlier morphological types.
%These central peaks may possibly be related to enhanced star formation
%activity.
We attribute part of this increase to the
effects of a steeply rising rotation curve at our relatively low
(0.2-1.2 kpc) spatial resolution in these galaxies. 
Thus, for these
galaxies, we measure the average velocity dispersion in the outer
disk excluding a central circular region. 
The exclusion aperture was chosen by examining
the individual velocity dispersion maps to identify the smallest
aperture that would exclude the region of most obviously enhanced
dispersion. We found that a diameter of
5.3 kpc (9 pixels  or 65$^{\prime\prime}$ at the distance of Virgo)
worked well for all five galaxies in the Virgo cluster and that
a similar physical region was appropriate for the other galaxies in
our sample. However, the maps of  NGC 4736 and NGC 4826 contain no
data at larger radii, and so for these two galaxies we quote disk
velocity dispersions excluding only a central 9 pixel diameter
region. The disk velocity dispersions of these two galaxies are likely
biased to be somewhat higher than those of the other galaxies in our
sample and thus we do not include them in calculating the average
outer disk velocity dispersion.

The average observed
velocity
dispersions in the molecular gas measured for the 12 galaxies in our 
sample are given in Table~\ref{tbl-props}. 
These velocity dispersions are consistent with previous
measurements for the Milky Way \citep{sb89},
M33, \citep{ws90} and other nearby galaxies \citep{cb97,w02}.
The observed velocity dispersion in
the molecular gas disk ranges from a low of 4.1 km s$^{-1}$ in NGC 628 to a
high of 20.1 km s$^{-1}$ in NGC 4501. We exclude NGC 4736 and NGC 4826
from our averages because we can only trace the velocity dispersions
in the inner region where the profile is still rising steeply. We
further exclude NGC 4501 because its very high velocity dispersion and
evidence for double peaked line profiles in the north-west portion of
the disk suggest that the molecular gas has been affected by the same
ram pressure effects seen in HI \citep{v08}.
The average observed value for the remaining 9 galaxies is $7.1\pm 0.9$
km s$^{-1}$  
(standard deviation 2.6 km s$^{-1}$), 50\%
smaller than the average of $11\pm 3$ km s$^{-1}$
measured for the atomic gas in the outer disks of spiral galaxies
\citep{l08}. If we compare the atomic and molecular velocity
dispersions in the same region of the disk (Table~\ref{tbl-props}),
the observed velocity dispersion in 
the molecular gas is on average $\sim 2$ times smaller than the dispersion in
the atomic gas. 

\subsection{The effect of cloud internal velocity dispersions}\label{sec-cc}

The low values measured for the velocity
dispersion, particularly for NGC 628 and NGC 3938, lead us to
consider the effect 
of internal velocity dispersions of individual giant molecular clouds
on the observed velocity dispersion. 
A second complicating effect, that 
of an anisotropy between the vertical and in-plane motions of the GMCs, would
be expected to produce a correlation of velocity dispersion with
inclination, which is not seen in our data
(Figure~\ref{fig-v_vs_i}). Nevertheless, we discuss
the potential magnitude of this effect in Appendix B.

Individual giant molecular clouds have internal velocity dispersions
that are substantially larger than the thermal line-widths expected
for gas with physical temperatures of 10-30
K. In the study of \citet{s87}, individual GMCs show internal velocity
dispersions ranging from 1- 8 km s$^{-1}$. These internal velocity dispersions
are obtained from an intensity-weighted measure of the dispersion in
radial velocity within a single cloud and thus are directly comparable
(in technique) to the intensity-weighted dispersions averaged over the
galactic disk presented here. What we need is an estimate of the
contribution of the internal velocity dispersions of an ensemble of
GMCs with a range of masses to the observed velocity dispersion in the
disk. With this value, it is easy to show that the observed velocity
dispersion, $\sigma_v$, is given by
$\sigma_v^2 = \sigma_{c-c}^2 + \overline{\sigma_{int}^2}$ where 
$\overline{\sigma}_{int}$ is the mass-weighted average internal velocity
dispersion for our ensemble of clouds and $\sigma_{c-c}$ is the value
we are interested in measuring, that is, the cloud-cloud velocity dispersion.

It is well known that the cloud internal velocity dispersion, $\sigma_{int}$,
relates to the cloud mass as $M \propto \sigma_{int}^4$, as first
proposed by \citet{h84}.
Observationally,
\citet{s87} find that the internal velocity dispersion $\sigma_{int}$ (in
km s$^{-1}$) relates to the cloud mass $M$  (in M$_\odot$) as $M =
2000 \sigma_{int}^4$. If we assume a cloud mass function $dN/dm \propto
m^{-\alpha}$, then we can estimate the mass-weighted average
internal velocity dispersion for an ensemble of clouds between $M_{low}$ and
$M_{high}$.  Since column density is proportional to intensity for
the CO lines \citep{s88}, a mass-weighted average is also equivalent to an
intensity-weighted average. The mass-weighted square of the 
internal velocity dispersion is 
then $\int m \sigma_{int}^2 dN / \int m dN$ which, using the relation
between $\sigma_{int}$ and $M$ given by \citet{s87} and assuming $\alpha
\ne 2$, reduces to
$$\overline{\sigma_{int}^2} = {1 \over 44.7} { 2 -\alpha \over 2.5 - \alpha}
{{M_{high}^{2.5-\alpha} - M_{low}^{2.5-\alpha}} \over
{M_{high}^{2-\alpha} - M_{low}^{2-\alpha}}}$$
For $\alpha = 1.5$ \citep{sss85}, $M_{high}= 10^6$ M$_\odot$ and
$M_{low}= 10^4$ M$_\odot$ \citep[e.g., ][]{s87}, 
${\overline\sigma_{int}}= 3.5$ km s$^{-1}$. The exact result clearly
depends on the details of the model.

Interestingly, this estimate of the average cloud internal velocity
dispersion is comparable to the value of 4.1 km s$^{-1}$ observed in
the disk of NGC 628.
The average CO $J$=3-2 intensity in NGC 628 of
0.8 K km s$^{-1}$ is the lowest intensity in our sample and 
translates into an average mass per JCMT beam of
only $4\times 10^6$ M$_\odot$. For a GMC mass function with slope
$\alpha=1.5$, this surface density translates into an average of just
10 clouds with $10^5 < M < 10^6$ M$_\odot$ per beam. This calculation
suggests that, although the number of massive ($> 10^5$ M$_\odot$)
GMCs per beam may be subject to small number statistics in NGC 628,
their effect on the velocity dispersion should  be observable on
average if such large clouds are present in NGC 628.
This analysis suggests that NGC 628 is likely
deficient the largest molecular clouds; if we instead limit the
maximum cloud mass to $10^5$ M$_\odot$, then the mass-weighted average
internal velocity dispersion reduces to ${\overline\sigma}_{int}= 2.2$
km s$^{-1}$. The larger average integrated intensities and observed
velocity dispersions suggests that the molecular cloud mass spectrum
is likely to be fully populated in the remaining galaxies in our
sample. Thus, in correcting the observed
velocity dispersions for the effects of internal cloud velocity
dispersions, we use a value for $\overline{\sigma}_{int}$ of 2.2 km
  s$^{-1}$ for NGC 628 and 3.5 km s$^{-1}$ for the rest
 of the sample. The corrected values for the cloud-cloud velocity
dispersion are given in Table~\ref{tbl-props} along with 
the uncertainty in the mean cloud-cloud
velocity dispersion for each galaxy. The uncertainty in the mean
 has been calculated
assuming that the errors are normally distributed, so that the
uncertainty in the mean is given by the 
standard deviation divided by $\sqrt{N}$, where $N$ is the number of
measurements for a given galaxy.

\subsection{The true vertical velocity dispersion of the molecular gas}

Based on the discussion in the
previous section, the observed velocity dispersion includes both a
contribution from the internal velocity dispersion of individual giant
molecular clouds as well as a contribution from the cloud-cloud
velocity dispersion. It is this second quantity, 
the cloud-cloud velocity dispersion, which is of interest for determining the
molecular gas scale height, analysing the stability of the gas disk,
etc. The small observed velocity dispersions imply that the correction
for the internal velocity dispersion of the clouds is not
negligible. Correcting for the average internal cloud velocity
dispersion as discussed in \S\ref{sec-cc} gives measures of the
cloud-cloud velocity dispersions which range from 2.7 to 19.8 km
s$^{-1}$. Again excluding NGC 4501, NGC 4736, and NGC 4826, we obtain 
an average value of $6.1 \pm 1.0$ km s$^{-1}$ (standard
deviation 2.9 km s$^{-1}$). 

\section{Discussion}\label{sec-disc}

\subsection{Molecular and atomic gas velocity dispersions}

The average cloud-cloud
velocity dispersion in the molecular gas in this group of  spiral galaxies 
of $6.1 \pm 1.0$ km s$^{-1}$ is slightly
smaller than previous estimates for external galaxies which did not
take the effect of internal cloud dispersion into account
\citep{cb97,w02}, but is 
comparable to recent measurements for the cloud-cloud dispersion of
massive clouds in the Galaxy 
\citep{sl06} and in M33 \citep{ws90}. 
%Our analysis has demonstrated
%that it is important to take into account the effect of the internal
%velocity dispersion of individual clouds on the observed velocity
%dispersion even when individual clouds are not resolved in the data. 

\subsubsection{Effects of CO optical depth}

An additional possible concern in interpreting the CO data is the high
optical depth of the CO line. Since we are interested in the
cloud-cloud velocity dispersion, it is important to check whether the
surface density of the molecular gas is sufficiently high that
shielding of one cloud by another at a similar velocity could affect
our measurements. For all but one of the galaxies in our sample, the
projected surface density  $N_H$ is substantially smaller than the
typical surface density of individual giant molecular clouds
\citep[$N_H =1.5 \pm 0.3 \times 10^{22}$ cm$^{-1}$, ][]{m89}, and thus
cloud-cloud shielding is unlikely to be a problem. Only in NGC 4736,
where the surface density may be as high as $3 \times 10^{22}$
cm$^{-2}$ \citep[assuming a CO J=3-2/J=1-0 line ratio of 0.3 and 
a CO to H$_2$ conversion factor of
$2\times 10^{20}$ cm$^{-2}$ (K km s$^{-1}$)$^{-1}$][]{s88}, is it
possible the clouds may shield one another to some 
degree. However, the actual surface density depends sensitively on
the value that we assume for the CO J=3-2/J=1-0 line ratio;
if we adopt a value of 0.6 which is more appropriate for the dense
gas, rather than the value of 0.3 found in a direct comparison of CO
J=3-2 and J=1-0 maps of NGLS galaxies \citep{w09,w10}, then
the projected disk-averaged column density is similar to that of
individual GMCs. Thus, we conclude that we can ignore the possibility
of cloud-cloud shielding in the disks of the galaxies in this sample.

\subsubsection{Observed HI velocity dispersions}

\citet{cb97}
showed that the molecular and atomic components
in NGC~628 and NGC~3938 had vertical velocity dispersions that are
similar to each other and virtually constant with radius. As a consequence,
the authors suggested that the two layers had similar scale heights and that
 HI and H$_2$ simply represent different phases of the same
dynamical component.  It is clear, however, from observations of 
our own Milky Way as well as edge-on galaxies that 
the molecular disk is 
significantly thinner than the HI disk.
Our observed velocity dispersions are comparable to
those found by \citet{cb97}, although
we have shown that
it is necessary to 
correct for internal velocity dispersion to determine the
cloud-cloud velocity dispersion that is relevant for scale height
determinations.  For HI, 
on the other hand, we know that its 
stable phases occur in both hot diffuse gas as well as
colder clouds (the warm neutral medium and cold neutral medium,
respectively). Although discrete HI clouds 
are identifiable in our Galaxy
\citep[see][]{l01,s06},
the clouds are smaller and warmer
than molecular clouds. Because of the admixture of cool and warm HI
components over the size scale of the beam in external galaxies,
 similar corrections need not be applied to the HI
velocity dispersions. Also, since the HI is a diffuse collisional
medium, its velocity dispersion is assumed to be isotropic
\citep[e.g.][]{m95}.
We have therefore
compared the observed HI dispersion directly with our cloud-cloud
velocity dispersion derived from the CO data.

\subsubsection{Comparing CO and HI velocity dispersions}

The published HI velocity dispersion map for NGC 4501
\citep{v08} shows 
velocity dispersions in the range of 10-20 km s$^{-1}$ across the extended
disk, comparable to the cloud-cloud velocity
dispersion (Table~\ref{tbl-props}) over the same area.
High resolution HI data have been
published for NGC 4321 \citep{k93} and NGC 4254 \citep{p93}, but the
velocity dispersion maps are not presented; we could find no published
high resolution HI data for NGC 4303 and NGC 4535.
An older measurement of the HI velocity
dispersion of 10 km s$^{-1}$ for NGC 3938 \citep{sk82} is roughly
3 times larger than our derived cloud-cloud velocity dispersion.

We have compared the velocity dispersions in the molecular and atomic gas
measured over the same regions for six galaxies in our sample for
which maps of the atomic gas velocity dispersion are available from
the THINGS sample \citep{w08}. On average, 
the cloud-cloud velocity dispersion determined from CO data is roughly
half that of the atomic gas. 
The exception to this trend is
NGC 4826, for which the cloud-cloud velocity dispersion is a factor of 5 times
smaller than the the HI velocity dispersion. However, this galaxy has quite
a steep radial gradient in the observed velocity dispersion
(Figure~\ref{fig-radial}) and so the average disk value depends 
sensitively on what aperture is used.
Overall, these results imply that the
scale height of the molecular gas is roughly half that of the
atomic gas. This estimate of the ratio of the scale heights is roughly 
consistent with the estimate of the relative HI 
and CO scale heights in the Milky Way as discussed in \citet{cb97}.  
We should note, however, that narrow (4-5 km s$^{-1}$) HI line
components have been seen in a few 
Local Group dwarf galaxies observed with high spatial resolution
\citep{l96,l97,d06}. These narrow lines, which may represent atomic
gas cooling to form the precursors of GMCs, are typically
superimposed on a broader underlying line which likely represents the
overall warm neutral atomic medium.

\subsection{Correlations with other global properties}

The infrared luminosity is a useful tracer of the star formation
activity within a galaxy \citep{k98}. Enhanced star formation might be
expected to 
increase the velocity dispersion in the interstellar medium via
stellar winds and supernova explosions. Indeed, \citet{t09} find
evidence that supernovae linked to recent star formation are important
for maintaining the velocity dispersion in the atomic gas.
We find
a statistically significant correlation (95\% confidence level) of 
both the observed and the cloud-cloud velocity dispersion with 
infrared luminosity \citep{s03} for the 9 galaxies in our sample.
Thus, like the atomic gas, there is some
evidence for star formation activity enhancing the velocity dispersion
in the dense molecular gas.
We also find a statistically significant correlation
(95\% confidence level) of both the observed and the cloud-cloud
velocity dispersion with 
the absolute K-band and absolute B-band
magnitude \citep{j03}.
Assuming the K-band magnitude traces the total stellar mass, this
correlation suggests that the cloud-cloud velocity dispersion is
enhanced in more masssive galaxies. For all these correlations,
however, it is important to keep in mind that we have compared the
average velocity dispersion in the disk (excluding the central
regions) with a global luminosity which would include the contribution
from the central regions.

There also appears to be a  dependence in our sample of the
velocity dispersion on galaxy morphology. We divided our sample into
early (type ab-bc) and late (type c-cd) spirals and calculated the
average velocity dispersion for each sub-sample. The average
observed dispersion for 6 late-type spirals is
$5.8 \pm
0.3$  km s$^{-1}$, while the value for 3 early-type spirals is  $9.7\pm
0.6$ km s$^{-1}$. 
These two values differ at the 6
sigma level. Of course, this dependence on morphology could be tracing
some other variable in our sample, such as mass or star formation activity.
In fact, the average infrared luminosities differ by a
factor of three between the early- and late-type samples, while the
K-band luminosity differs by a factor of four. Thus, it seems most
likely that the dependence on morphology is driven by differences in the
average star formation rates and masses between the early and late
type samples.

\subsection{Disk stability and star formation laws}

The velocity dispersion of the interstellar medium is an important
parameter in understanding the stability of galactic disks
against gravitational collapse.
\citet{j84a,j84b} showed that a rotating disk composed of two fluids with
different surface densities and velocity dispersions can be unstable
to perturbations even if each fluid disk is stable on its own. They
also showed that a galaxy must be treated as a two fluid system when the
colder gaseous component comprises as little as 10\% of the total mass
of the system. Treating the stellar component as a collisional fluid,
they found that the critical velocity dispersion required for
stability in this higher dispersion system is larger in the presence
of a second cold fluid than if the stellar system is treated in isolation.
\citet{r01} considered the case of a single fluid and multiple
collisionless (stellar) components, but found that the stability
conditions were quite similar in the two fluid case whether or not the
stellar component was treated as collisionless.

\citet{y07} applied the approach of \citet{r01} to compare the
location of star forming regions in the Large Magellanic Cloud (LMC)
with the disk stability criterion. They adopted a gas velocity
dispersion of 5 km s$^{-1}$, very similar to the value found in our
sample of spiral galaxies. They found that there was a much better
spatial correlation of star forming regions with regions where the
disk was unstable when instability was evaluated using the
contribution of both the gas and the stars. \citet{l08} applied the
two collisional fluid equation of \citet{r01} to a number of galaxies
in the THINGS sample \citep{w08}. They adopted a  gas velocity
dispersion of 11 km s$^{-1}$ from measurements of the HI
component. Similarly to \citet{y07}, they found that the $Q$ values
were smaller when both the stars and the gas were included in the
analysis; however, they found that the disks were globally stable
against large scale perturbations.

The results for the molecular velocity dispersion presented here imply
that the interstellar medium in spiral galaxies is itself a
multi-fluid system, with the molecular gas disk being dynamically
significantly colder than the atomic gas disk. Since
previous analyses have found that the colder of the two components has
the larger effect on the disk stability \citep{j84a}, it seems likely
that, if we were to use one phase of the ISM in our stability
analysis, it should be the dynamically coldest phase, which is the
molecular gas. This would be especially true for the inner regions of
those spiral disks where the ISM mass is predominantly H$_2$.
% next sentence checked and revised August 10
The Toomre $Q$ parameter for the combined system
tends to decrease 
as the velocity dispersion of the coldest (gas) component decreases and 
as the gas mass fraction ($\Sigma_{gas}/\Sigma_*$) increases
\citep{j84b,r01}. 
This effect suggests that the $Q$ values derived by
\citet{l08} are likely to be significantly larger than the values that
would be derived if the molecular gas values for surface density and
velocity dispersion were used in the analysis, a possibility that
 was discussed in \citet{l08} as well. It would be informative
to repeat the stability analysis for the galaxies in common between
our sample and the THINGS sample to see if the disks remain globally
stable as judged by the $Q$ parameter when the cloud-cloud velocity
dispersion of the molecular gas is included. We plan to investigate
this issue further in a future paper.

The need to treat the ISM itself as a multi-fluid system suggests that
we may not be able to separate the discussion of the instabilities in
the atomic and molecular phases. For example, we might imagine that the
$Q$ value for the atomic gas governs the formation of molecular clouds
from the HI disk, while the star formation from those clouds would
then be governed by the $Q$ value for the molecular disk. However, the
two fluid analysis by \citet{j84a} and \citet{r01} implies that the
$Q$ value and hence the stability properties of the atomic disk are
affected by the presence of the dynamically colder
molecular disk. In particular, an atomic gas disk with a stable value
of $Q$ when considered in isolation might be found to be unstable when
the presence of a relatively small quantity of molecular gas is
included. Of course, a complete analysis would include the properties
of the 
stellar disk \citep[perhaps multiple components, as in ][]{r01} in
addition to the two fluid components of the ISM. Such an analysis is
beyond the scope of this paper.

\subsection{Implications for turbulent support of the ISM}

The scale height of a gaseous component in hydrostatic equilibrium
depends on both gravitational and pressure gradients. 
Gravitational gradients originate in the disk as well as a dark
matter halo potential.  The pressure will 
include internal thermal pressure, turbulent
pressure and, if important, pressure due to cosmic rays and magnetic
fields.
At the radius of the Sun in the Milky Way, the magnetic and cosmic ray
pressures supply about half of the total pressure and the thermal and
turbulent pressures supply the remaining half (Hanasz \& Lesch 2004).
At the temperatures of the molecular components, turbulent
pressure dominates over thermal pressure within and between clouds
\citep[e.g.][]{b03}.
In the HI, thermal pressure likely dominates
in the Warm Neutral Medium whereas both may be important for the Cold
Neutral Medium
\citep{ht03}.
Measured velocity dispersions can only probe the dynamical components
(thermal and turbulent)
and therefore can only predict the scale height of the gas provided
that the dynamical components dominate over the magnetic and cosmic ray
pressures 
%(a situation that 
%is more likely to occur at larger galactocentric radii) 
and provided
that the disk ISM is in pressure equilibrium.
% 14/10/2010 Elias wants this sentence clarified; emailed Judith
% who said to remove the comment in brackets

Star formation activity is typically associated with heating sources
and sources of turbulence through supernovae and stellar winds.
%Roughly half the galaxies in 
Most of the galaxies in
our sample show a radial decline in the CO velocity dispersion
(Figure~\ref{fig-radplot_all}).
The velocity dispersion in HI
is also a declining function of radius \citep{t09}, although
HI disks retain significant velocity dispersions outside
any star forming disk 
\citep[e.g.][]{sb99,t09}.
\citet{t09} found a good correlation between the kinetic energy of the
HI and the star formation rate and suggest that supernovae are likely
sufficient to maintain the HI velocity dispersion in regions with
significant star formation.
The correlation of the cloud-cloud velocity dispersion
with the star formation rate as traced by the infrared luminosity
suggests that star formation may also be the dominant source of turbulence
in the H$_2$-rich parts of spiral disks.

However, some galaxies do show a lack of correlation between
turbulence in the atomic gas and
star formation as a function of radius.
\citep[e.g.][]{d90,zb99,pr07}.
%Roughly half of our galaxies show a roughly
The three lowest luminosity galaxies in our sample show a roughly
constant cloud-cloud velocity dispersion
as a function of radius \citep[see also][]{cb97}
These results suggest that non-thermal energy input in the form
of turbulence that is unrelated to star formation may be important for
understanding the vertical velocity dispersions in disks.
Magnetic instabilities may be the most promising source of turbulence.
Parker instabilities \citep{p66}
are well-known, but seem to require
cosmic rays from supernovae as triggers to be most effective 
\citep{hl00}. Thus, Parker instabilities cannot
account for the observed dispersions seen outside of the active
star forming disk, but could be important for the inner molecular
disks where star formation is occurring.  Another possibility is the
Balbus-Hawley  
(or magneto-rotational)
instability \citep{bh91,bh98},
or other instabilities related to magnetic stresses 
\citep[e.g.][]{sb99}.
\citet{t09} suggest that a combination of thermal broadening and
magneto-rotational instabilities can account for the HI velocity
dispersion beyond $r_{25}$.
Simulations of the magneto-rotational instability by 
\citet{po07}
for the two-phase HI component have shown that the turbulent velocity
amplitude varies as $\propto n^{-0.77}$, where $n$ is the mean density
of the component.  If this turbulence is dominant in both molecular
and atomic gas, then the difference in velocity dispersion
between the two components may simply relate to their respective 
densities. 

\subsection{Disk galaxies at high redshift}

Observations of the molecular gas content in galaxies at redshifts $z
= 1-3$ are often made using the CO J=3-2 line, which at these
redshifts falls into the standard millimetre observing windows. Most
high-resolution detections of high redshift galaxies are unusually
luminous systems, such as submillimetre galaxies \citep{t08} or
quasars \citep{copp08}. Although many of these high redshift
detections appear as simple point sources at current resolution
limits, there are a few systems for which we can begin to resolve the
velocity structure. Even more interesting from the point of view of
this paper are the observations of more normal galaxies at redshifts
1-1.5, which are
becoming available \citep{d08,d09,d10,tacc10}. These galaxies are found to
be quite extended ($\sim 10$ kpc), significantly more so than the
quasars and submillimeter galaxies \citep{d10,tacc10}
Given the
sensitivity and wide field of view of our observations, it is
interesting to examine how our galaxies would appear if placed at a
redshift of 1.

\begin{figure}
% (a) is identical to Figure 2(a)
\includegraphics[width=84mm]{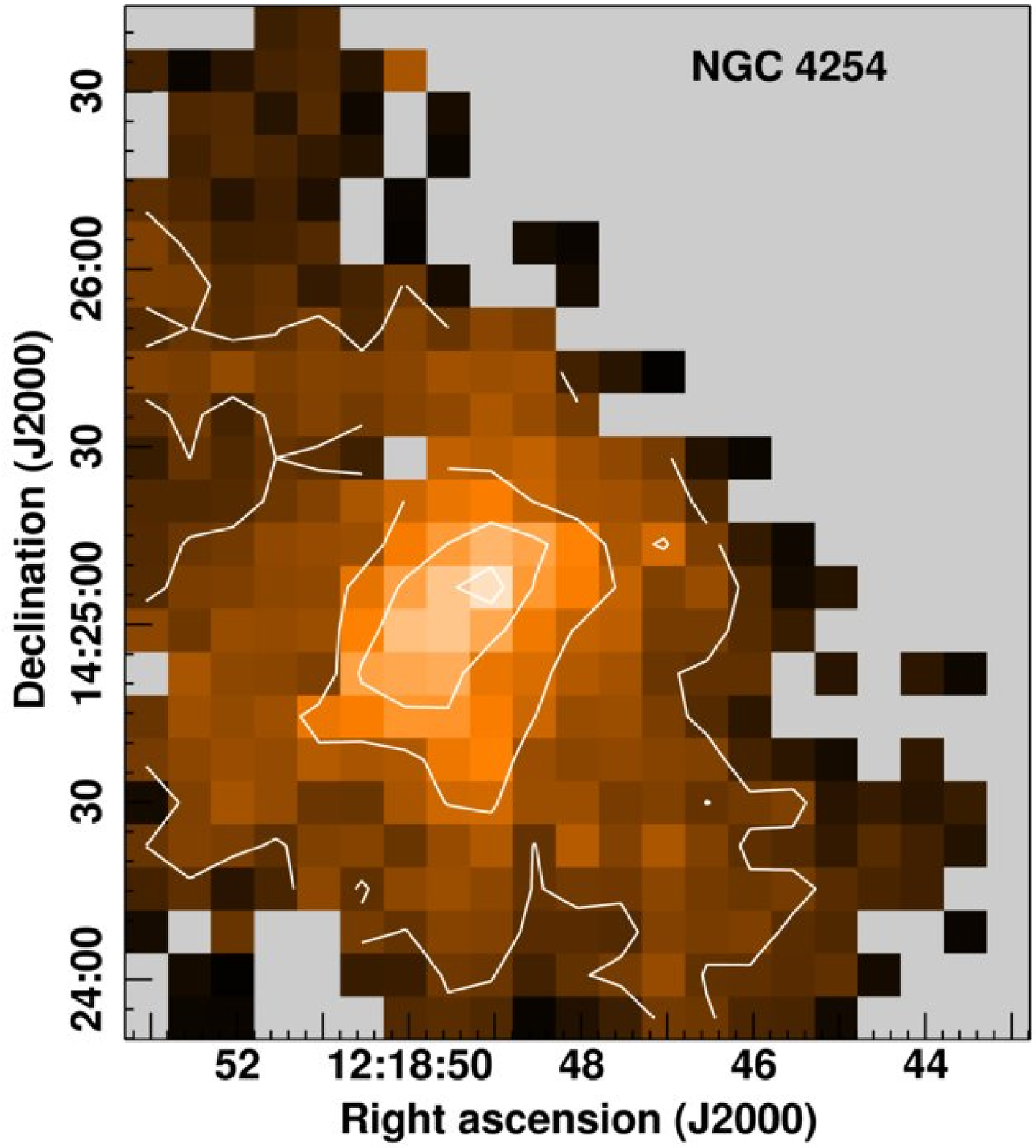}
\includegraphics[width=84mm]{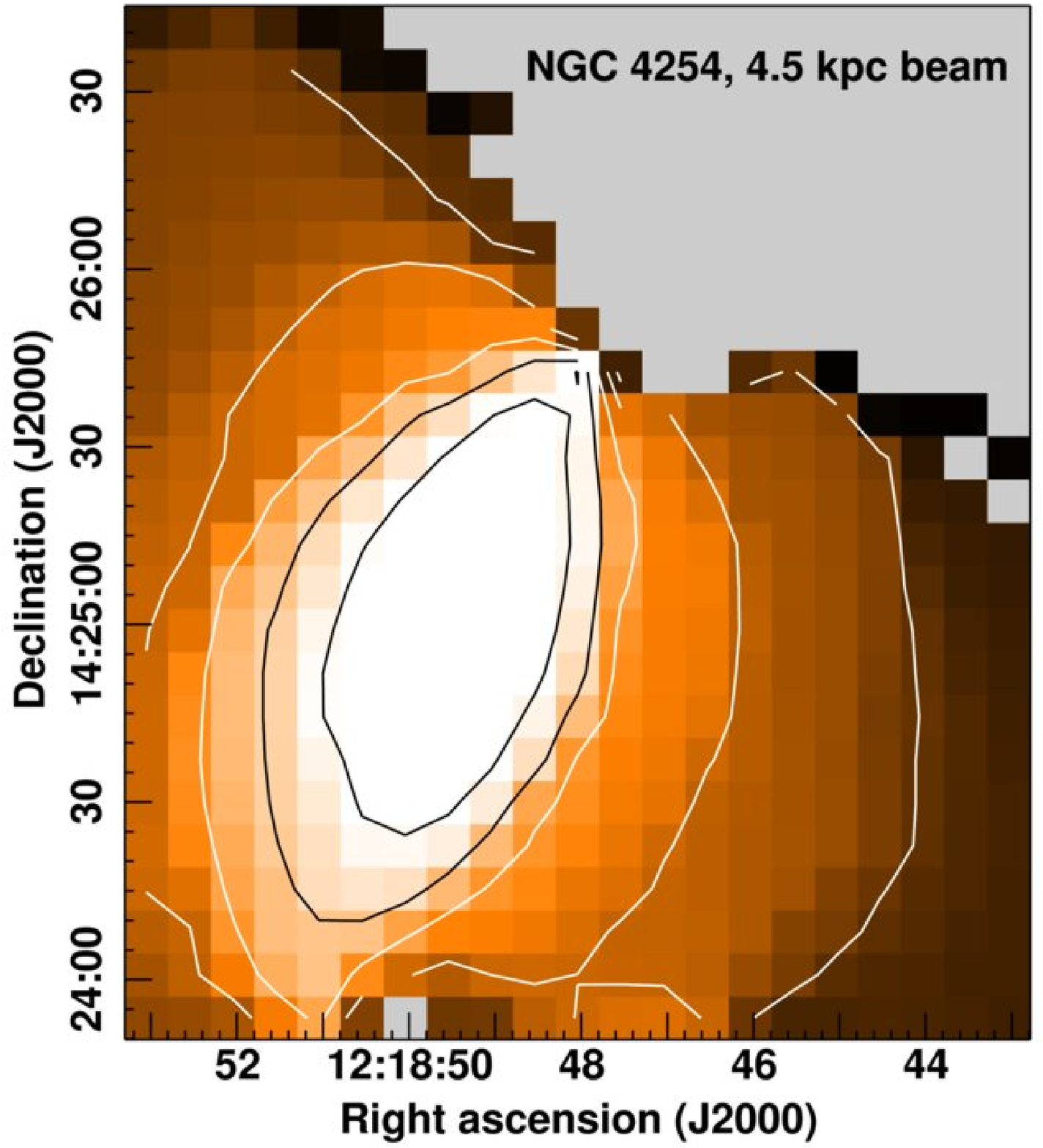}
\caption{(a) CO $J$=3-2 velocity dispersion for NGC 4254 at the native
  resolution of the JCMT data (14.5$^{\prime\prime}$ or 1.15 kpc at a
  distance of 16.7 Mpc). Colour scale
  runs from 0 to 60 km s$^{-1}$ and the contours are 10, 20, 30, 40,
  50 km s$^{-1}$. 
(b) CO $J$=3-2 velocity dispersion for NGC 4254 smoothed to a
resolution of 4.5 kpc (56$^{\prime\prime}$) to match the resolution of
observations of disk galaxies at $z \sim 1$ of \citet{tacc10}. Colour
scale and contours are the same as in (a).
\label{fig-highz}}
\end{figure}

We have processed a subset of our galaxies to see what changes are
produced in an analysis of the velocity dispersion when the spatial
resolution of the data is degraded. For three of the brightest
galaxies in the Virgo Cluster (NGC 4254, NGC 4321, and NGC 4303), we
convolved the baseline-subtracted data cube with a 54$^{\prime\prime}$
 gaussian to achieve an effective beam of 56$^{\prime\prime}$ or 4.5
 kpc at a distance of 16.7 Mpc. We then constructed moment maps from
 the smoothed data cube using the methods described in
 \S~\ref{sec-obs}.
A comparison of the resulting velocity dispersion maps
 for NGC 4254 is shown in Figure~\ref{fig-highz}. Depending on how
 much of the central high dispersion regions is excluded from the
 analysis, the average velocity dispersion in the outer disk of the
 low resolution image is on average twice as large as
the value
 in the high resolution image. Similar increases of about a factor of
 two in the average velocity dispersion in the disk are seen for NGC
 4321 and NGC 4303. 
The images also suggest that a plot of velocity
 dispersion as a function of radius would yield higher values at a
 given radius in the low resolution image. These effects of resolution
 will need to be taken into account as we begin to accumulate data on
 the dense molecular gas properties in galaxies at high redshift.

The average values of 20-30 km s$^{-1}$ seen at
 low resolution in the disks of these three galaxies are quite similar
 to the value  of $\sim$20 km s$^{-1}$ found in EGS13035123 by \citet{tacc10}.
EGS13035123 has 10-20 times the star formation rate, H$_2$ gas mass,
and stellar mass compared to NGC 4254
\citep{k03,w09,kranz03}. However, EGS13035123 is also roughly a factor of three
  larger in its CO radius than NGC 4254 \citep{w09}, which implies
  that the gas surface density, stellar surface density, and star
  formation rate surface density inside the molecular disks of
these two galaxies are quite similar. 
Thus, the similar
velocity dispersions between the two galaxies are likely related to
the similar mass surface densities in the disk.
The two galaxies also have quite
similar molecular gas fractions of $\sim 0.25$ \citep{tacc10,w09,kranz03}.
This comparison supports a picture where the high star formation rates
seen at $z=1$ may be at least partly due to the presence of physically
larger molecular gas disks at this epoch.

% why did Linda use such old references and a dwarf galaxy survey for
% her gas mass fraction of z=0 spirals??

%This lack of major evolution is
%consistent with other recent studies of disk galaxies. For example,
%\citet{f07} find the number and total stellar mass in blue
%galaxies is roughly constant from $z=1$ to the present. In addition,
%\citet{k07} find no evidence for a significant variation in the
%Tully-Fisher relationship over the same redshift range.

\section{Conclusions}\label{sec-concl}

We have used large-area high velocity resolution CO $J$=3-2
observations of 12 nearby galaxies to study the vertical velocity
dispersion in the dense molecular gas. 
Three of the galaxies show a roughly constant velocity dispersion as a
function of radius, while the other nine galaxies have a central peak
followed by a fall-off with radius to typically $0.2-0.4
r_{25}$. 
Flat velocity dispersion profiles are seen only in some of the
late-type spiral galaxies in our sample, and those with flat profiles
have the lowest mass.
The observed values of the
velocity dispersion range from 4.1 km s$^{-1}$  to 20.1 km
s$^{-1}$. These velocity 
dispersions are comparable to the internal velocity dispersions of
individual giant molecular clouds in our own and other galaxies
\citep{s87,ws90}. Correcting for these internal velocity dispersions
yields an average cloud-cloud velocity dispersion of $6.1 \pm 1.0$ km
s$^{-1}$ measured over 9 galaxies with good radial profiles. 
This cloud-cloud velocity dispersion is comparable to recent
measurements in our 
own Galaxy \citep{sl05,sl06} and M33 \citep{ws90}.

A direct comparison with the high resolution HI maps from the THINGS
survey \citep{w08} in six galaxies show that the cloud-cloud velocity
dispersion is on average twice as small as the velocity dispersion
of the atomic gas, 
which implies a much smaller scale height for the molecular gas. 
Theoretical analyses of the stability of multi-component disks suggest
that the dynamically coldest component is the most important in
driving instability \citep{j84a,j84b,r01}. This analysis suggests that it is
the properties of the dense molecular gas, rather than the atomic gas,
that are the most important for determining whether galactic disks
are stable against gravitational collapse, especially where the mass of the
ISM is H$_2$ dominated. 

The cloud-cloud velocity dispersion is correlated at the 95\%
confidence level 
with both the far-infrared luminosity
and the K-band absolute magnitude. Thus, as for the atomic gas, we
find evidence that star formation activity (as traced by the infrared
luminosity) tends to increase the velocity dispersion in the dense
molecular gas. The correlation with K magnitude, which traces the
total stellar mass, implies that the cloud-cloud velocity dispersion
is also enhanced in more massive galaxies.

We have used our data to examine the apparent kinematical properties
of the molecular disk at a spatial resolution of 4.5 kpc chosen to
match the best resolution for galaxies at
redshifts 1-2 \citep{t08}. A degradation of the resolution from 1.2
kpc to
4.5 kpc results in an increase in the average velocity dispersion in
the outer disk by a factor of two. 
The average velocity dispersion of
NGC 4254 viewed with 4.5 kpc resolution is quite comparable to the
velocity dispersion of 20 km s$^{-1}$ seen in a normal galaxy at $z=1$
\citep{t09}. Both galaxies have comparable gas and stellar surface
densities, as well as star formation rate surface densities, which
suggests that the higher star formation rates seen at $z=1$ may be
partly attributed to the presence of physically larger molecular disks.
This analysis suggests that this data set can provide a
valuable local bench mark in understanding lower spatial resolution
observations of galaxies in the early universe.

\section*{Acknowledgments}

We thank the anonymous referee for a referee report which spurred us
to re-examine our data processing choices and resulted in a
significant improvement in the paper.
The James Clerk Maxwell Telescope is operated by The Joint Astronomy
Centre on behalf of the Science and Technology Facilities Council of
the United Kingdom, the Netherlands Organisation for Scientific
Research, and the National Research Council of Canada. The research of
J.I. and C.D.W. is supported by grants from NSERC
(Canada). A.U. has been supported through a Post Doctoral Research
Assistantship from 
the UK Science \& Technology Facilities Council.
Travel support for
B.E.W. and T.W. was supplied by the National Research Council (Canada).
We acknowledge the usage of the HyperLeda database
(http://leda.univ-lyon1.fr) and thank R.N. Henriksen for useful
discussions. This research has made use of the 
NASA/IPAC Extragalactic Database (NED) which is operated by the Jet
Propulsion Laboratory, California Institute of Technology, under
contract with the National Aeronautics and Space Administration. 

%{\it Facilities:} \facility{JCMT}.

\appendix
\section{Comparison with previous processing methods}

There are two different methods that have been used to measure
velocity dispersions in molecular gas in galaxies. \citet{cb97}
measured velocity dispersions in NGC 628 and NGC 3938 by fitting
gaussian profiles to individual spectra, while \citet{w02} used moment
2 maps of NGC 6946 to determine the average velocity dispersion. While
the moment 2 maps used in this paper are easier to compute and analyse
than fitting individual profiles to many hundreds of spectra, they can
potentially be subject to some systematic effects depending on choices
made in the processing. We investigate some of these effects here by
comparing our data and moment 2 maps for NGC 628 and NGC 3938 with the
results given by \citet{cb97} for the same galaxies.

We first investigated whether the measured velocity dispersion was
affected by the angular and spectral resolution of the data. The CO
J=1-0 data from \citet{cb97} have an angular resolution of
23$^{\prime\prime}$ and a spectral resolution of 2.6 km s$^{-1}$, while the
JCMT CO J=3-2 data have an angular resolution of
14.5$^{\prime\prime}$ and a spectral resolution of 0.43 km s$^{-1}$. We
produced three additional data cubes for each galaxy: one binned by 6
channels in velocity, one convolved to achieve a gaussian beam of
23$^{\prime\prime}$, and one that was both convolved and binned. From
these cubes we made moment 2 maps using  the method described in
\S~\ref{sec-obs} but using a signal-to-noise cutoff for the mask of
3.5 sigma so as to include only regions with relatively high signal to
noise. The average velocity dispersion measured in each map is given
in Table~\ref{tbl-a1}. These results clearly show that data with
poorer spectral or spatial resolution will give higher values for
the velocity dispersion than data with better spectral or spatial
resolution. The data also suggest that {\it spatial} resolution may be
a more important factor in increasing the measured velocity dispersion
than spectral resolution, as long as the
spectral resolution is sufficient to resolve the lines. Interestingly,
our average velocity dispersions for these two galaxies are only
slightly smaller than those measured by \citet{cb97} when our data are
smoothed spatially and spectrally to match the IRAM data.

\renewcommand{\thefootnote}{\alph{footnote}}
\begin{table}
 \centering
  \caption{Velocity dispersions measured with different spectral and
    spatial resolutions\label{tbl-a1}}
  \begin{tabular}{lcc}
  \hline
Data cube used\footnotemark[1] & NGC 3938 & NGC 628\\
\\
 \hline
Original & 3.7 & 3.1 \\
Binned to 2.6 km s$^{-1}$ & 4.8 & 3.9 \\
Convolved to $23^{\prime\prime}$ gaussian & 6.5 & 4.2 \\
Convolved and binned & 7.4 & 4.7 \\
\citet{cb97} & 9 & 6.5 \\
 \hline
\end{tabular}
\begin{tabular}{l}
\footnotemark[1] All values measured for the CO J=3-2 line in km
s$^{-1}$. \\
Velocity dispersion measured using a signal-to-noise cutoff of 3.5$\sigma$.
\end{tabular}
\end{table}

We next investigated the effect of the choice of the signal-to-noise
cutoff for the masks used in creating the moment map. For this
analysis, we used the data cubes that had been smoothed 
spatially to match the data from \citet{cb97} to give better
signal-to-noise in our maps, but which had no spectral smooting
applied. 
The results are given
in Table~\ref{tbl-a2}. It is clear that the measured velocity
dispersion decreases systematically has higher signal-to-noise cutoffs
are used in the mask.

\renewcommand{\thefootnote}{\alph{footnote}}
\begin{table}
 \centering
  \caption{Velocity dispersions measured with different
    signal-to-noise cutoffs for mask\label{tbl-a2}}
  \begin{tabular}{lcc}
  \hline
S/N cutoff used\footnotemark[1] & NGC 3938 & NGC 628\\
\\
 \hline
5$\sigma$ & 3.8 & 3.4 \\
4.5$\sigma$ & 4.8 & 3.7 \\
4.0$\sigma$ & 5.7 & 3.9 \\
3.5$\sigma$ & 6.5 & 4.2 \\
3.0$\sigma$ & 7.2 & 4.6 \\
2.5$\sigma$ & 9.1 & 5.3 \\
\citet{cb97} & 9 & 6.5 \\
 \hline
\end{tabular}
\begin{tabular}{l}
\footnotemark[1] All values measured for the CO J=3-2 line in km s$^{-1}$. \\
\end{tabular}
\end{table}

\begin{figure}
\includegraphics[angle=0,width=84mm]{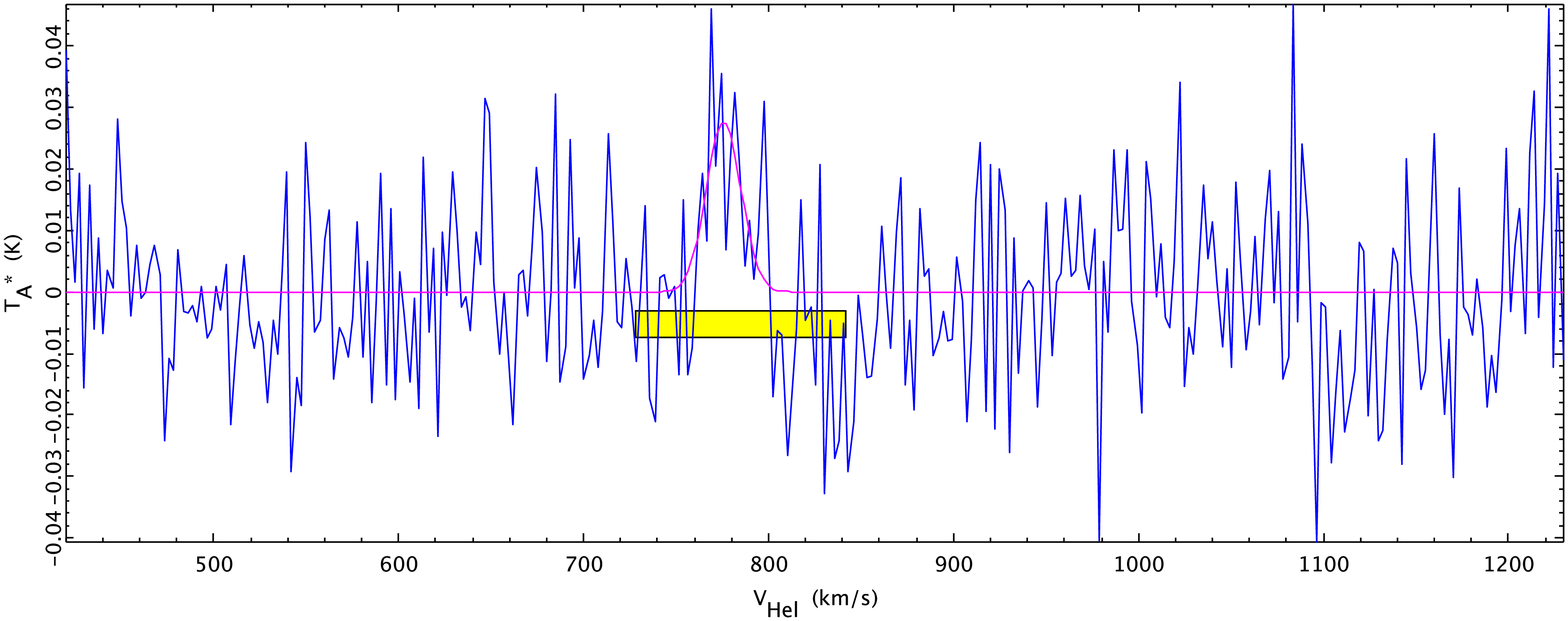}
\includegraphics[angle=0,width=84mm]{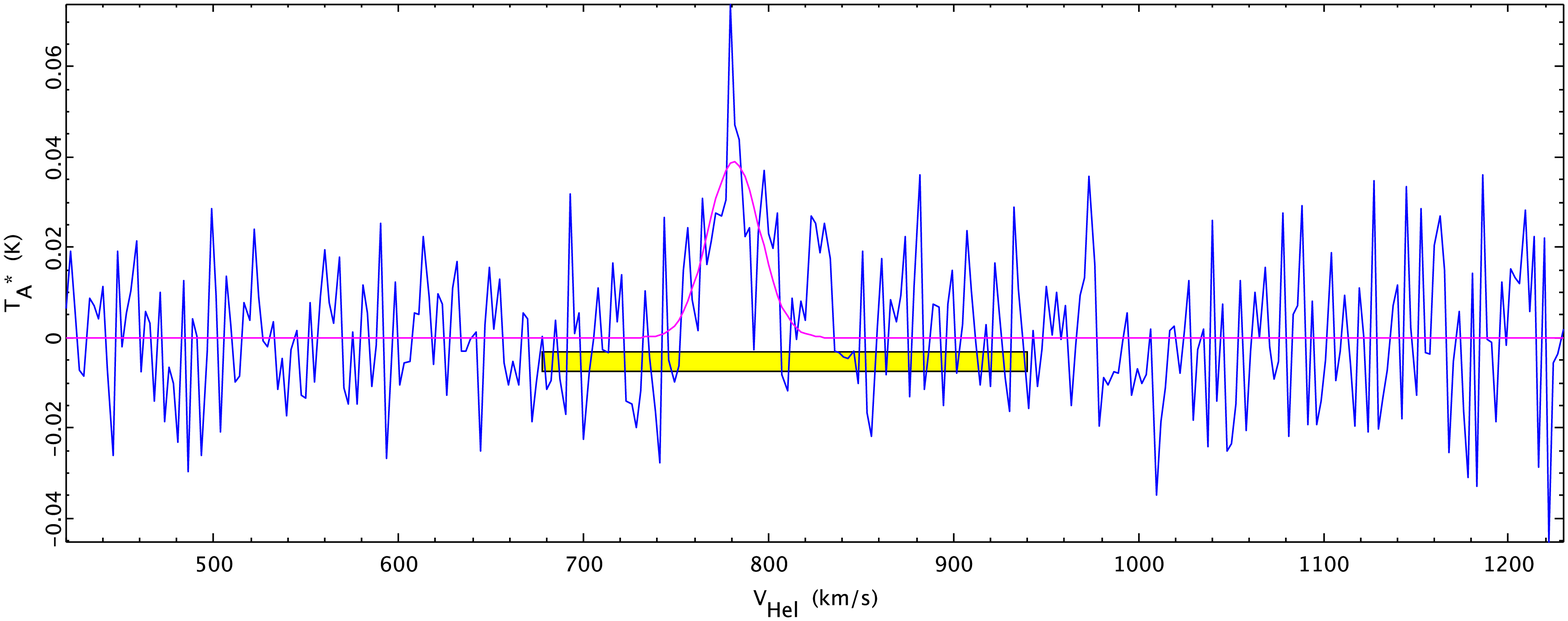}
\includegraphics[angle=0,width=84mm]{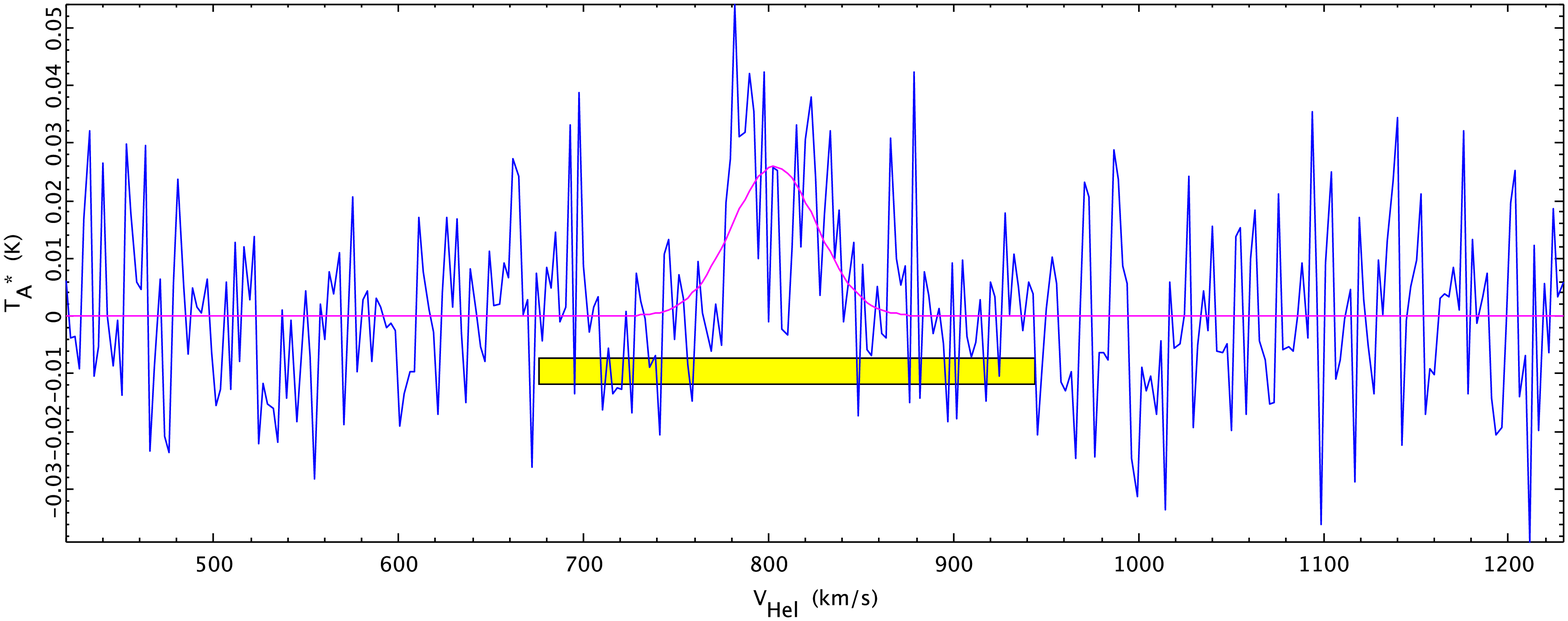}
\includegraphics[angle=0,width=84mm]{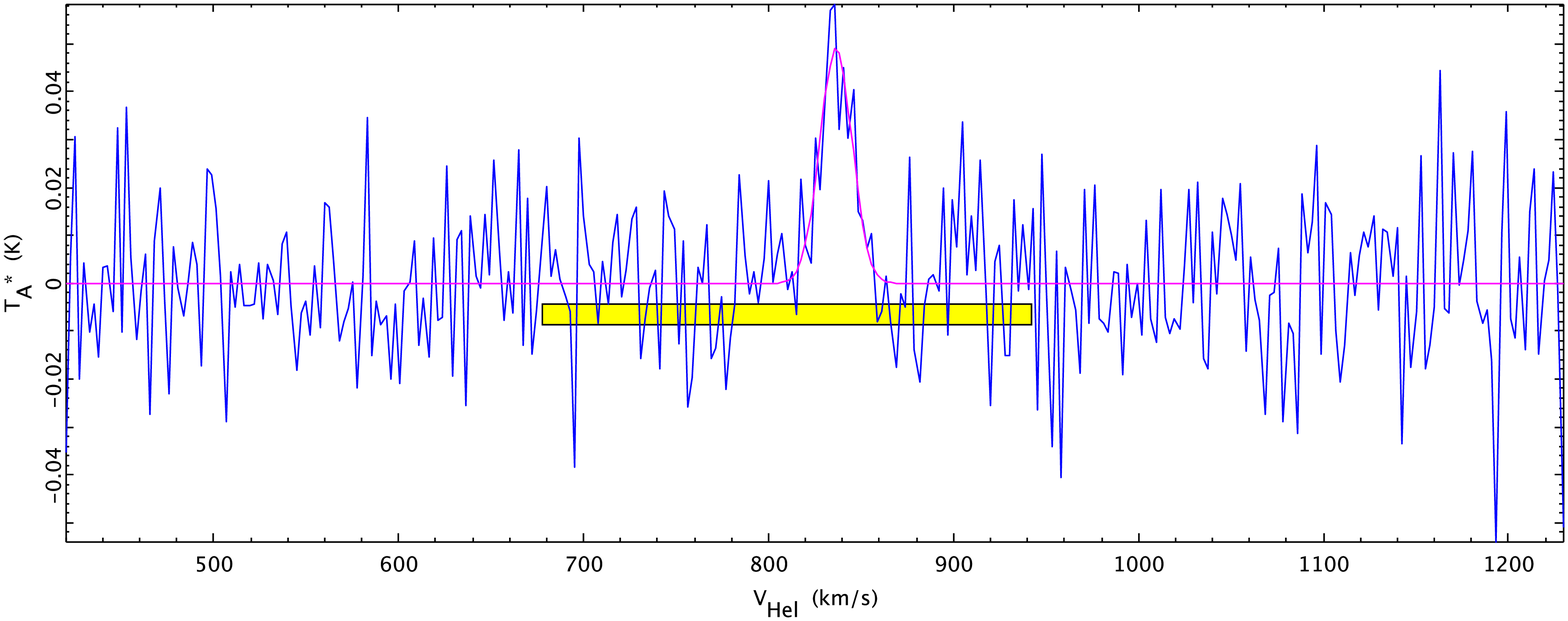}
\includegraphics[angle=0,width=84mm]{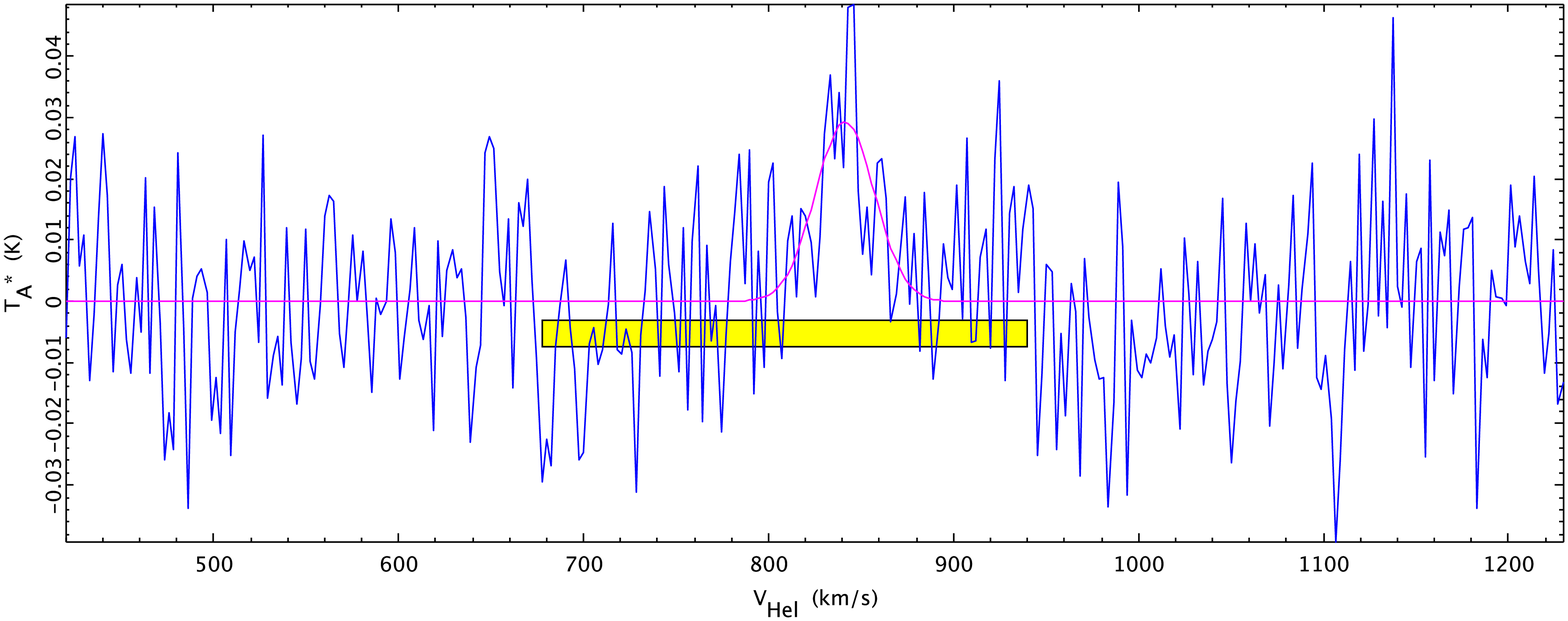}
\caption{Observed CO $J$=3-2 emission at five positions in NGC 3938.
The spectra have been convolved 
to a 23$^{\prime\prime}$ beam and
binned to 2.6 km s$^{-1}$ resolution. Spectra are spaced by
22$^{\prime\prime}$ along the declination axis ordered from north
(top) to south (bottom) and the central
spectrum corresponds to (11:52:49.6,44:07:17.7). Gaussian fits to each
spectrum are overlaid; the bar indicates the region used to obtain the fit.
\label{fig-spectra}}
\end{figure}

Finally, we performed gaussian fits to selected spectra from the data
cube of NGC 3938 that had been convolved to 23$^{\prime\prime}$ and
binned to 2.6 km s$^{-1}$ resolution (Figure~\ref{fig-spectra} and
Table~\ref{tbl-a3}. To mimic approximately the selection of spectra
shown in \citet{cb97}, we selected the spectrum with the highest
velocity dispersion in a map made with a 3.5$\sigma$ cutoff, and then
stepped away from this spectrum in steps of 3 pixels
(21.8$^{\prime\prime}$) in the north and south directions. We obtain
similar values for the velocity dispersions from fits to the unbinned
spectra; binned spectra are shown in Figure~\ref{fig-spectra} for
clarity. Although there is considerable scatter in the velocity
dispersions for individual spectra, the average value derived from
gaussian fits is 1.2 times larger (the standard deviation 0.4) than
the value derived from the moment 2 map.  

\renewcommand{\thefootnote}{\alph{footnote}}
\begin{table}
 \centering
  \caption{Comparison of velocity dispersion from gaussian fits and
    second moment maps for selected positions in NGC 3938\label{tbl-a3}}
  \begin{tabular}{lcc}
  \hline
Position\footnotemark[1] & Gaussian fit & Moment 2 map\footnotemark[2] \\
($^{\prime\prime}$,$^{\prime\prime}$) & (km s$^{-1}$) & (km s$^{-1}$) \\
\\
 \hline
(0,44) & 9.4$\pm$ 2.9 & 7.8 \\
(0,22) & 14.0$\pm$ 2.3 & 19.3 \\
(0,0) & 23.0$\pm$ 4.4 & 19.3 \\
(0,-22) & 8.7$\pm$ 1.3 & 8.4 \\
(0,-44) & 15.0$\pm$ 3.2 & 8.2 \\
 \hline
\end{tabular}
\begin{tabular}{l}
\footnotemark[1] All values measured for the CO J=3-2 line. The
spectra have been convolved \\ 
to a 23$^{\prime\prime}$ beam and
binned to 2.6 km s$^{-1}$ resolution. (0,0) corresponds to \\
(11:52:49.6,44:07:17.7 J2000).\\
\footnotemark[2] Measured from maps made using a 2.5$\sigma$ cutoff.
\end{tabular}
\end{table}

On the basis of this analysis, we opted to use a signal-to-noise
cutoff of 2.5$\sigma$ on our original data cubes, as this choice
seemed to give good agreement with the results from \citet{cb97} when
both data sets were matched in angular and frequency resolution.

\section{The possible effect of anisotropies in the velocity dispersion}
\label{sec-anisotropy}

\begin{figure}
\includegraphics[angle=-90,width=84mm]{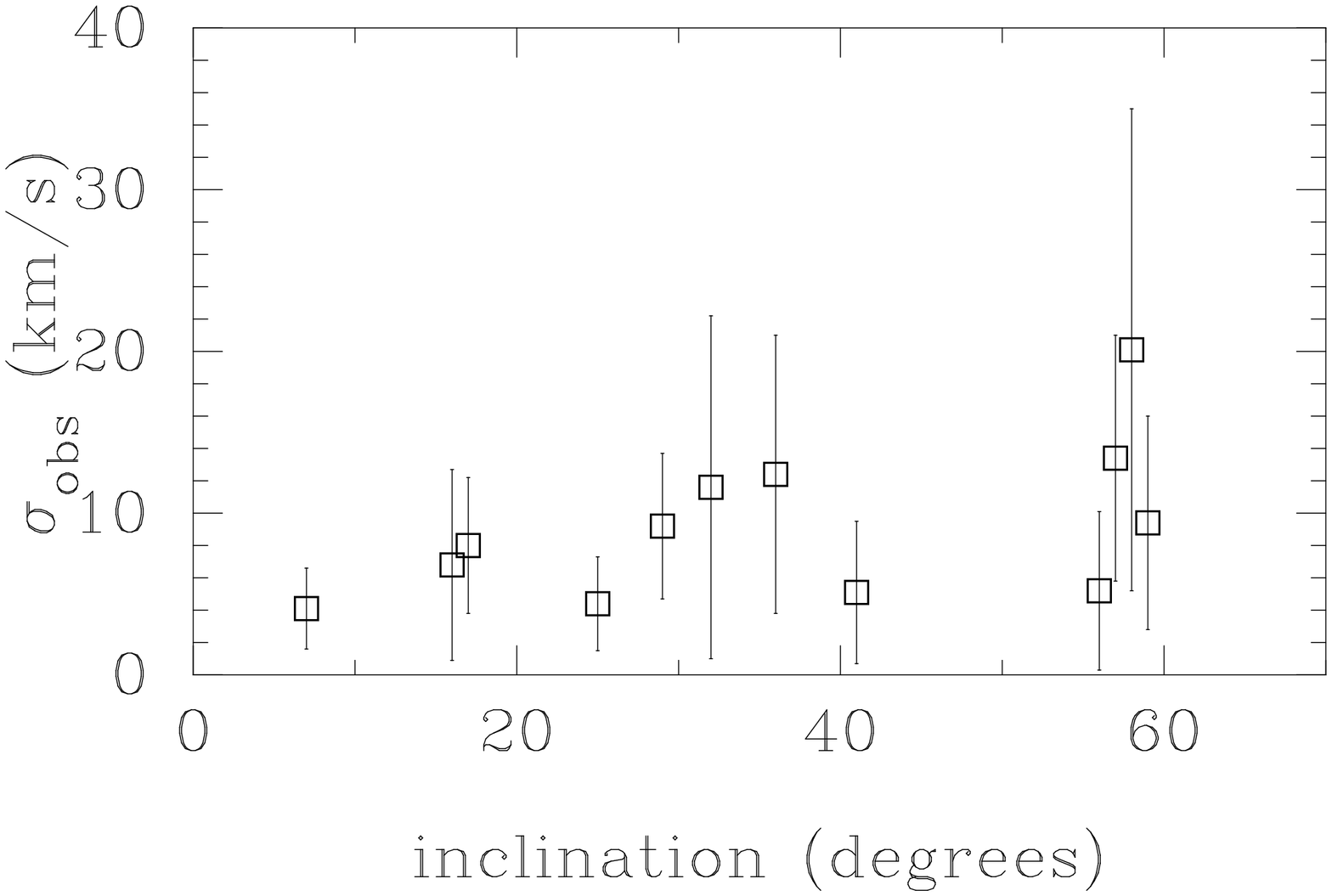}
\caption{Observed CO $J$=3-2 velocity dispersion as a function of
  galaxy inclination angle. Error bars show the standard deviation of
  the observed values within an individual galaxy.
\label{fig-v_vs_i}}
\end{figure}

There is no statistically significant correlation of velocity
dispersion in the 
molecular gas with inclination in our data
(Figure~\ref{fig-v_vs_i}). The average
observed velocity dispersion is 5.3 km s$^{-1}$ for the 4 galaxies with
inclinations $< 25^o$ and 8.1 km s$^{-1}$ for the 8 galaxies with
inclinations $> 25^o$ and these values agree within 2 sigma. 
We have also checked for any correlation of the velocity dispersion
normalized by each of the star formation rate and $D_{25}$ with
inclination and again find no correlation.

For an inclined galaxy, the observed  velocity
dispersion is a combination of the in-plane
($\sigma_r$,$\sigma_\theta$) and vertical
velocity ($\sigma_z$) dispersions. We will assume that $\sigma_r =
\sigma_\theta$ and will 
refer to the in-plane component of the
velocity dispersion  as $\sigma_r$ from here on
for simplicity. Thus, for a galaxy with inclination $i$, the observed
velocity dispersion is $\sigma_{obs} = \sqrt{\sigma_z^2 \cos^2 i 
+ \sigma_r^2 \sin^2 i}$.
If the radial and vertical velocity dispersions
of the molecular gas were isotropic, we would not expect any trend of
velocity dispersion with inclination. If, however, the velocity
ellipsoid  of the giant molecular clouds (which contain most of the
mass of the molecular ISM) is anisotropic, with $\sigma_r > \sigma_z$,
then the observed velocity dispersion 
would tend to increase with increasing
  inclination. 

Unfortunately, there is no direct information on the shape of the
velocity ellipsoid for giant molecular clouds. In theoretical models,
the cloud velocity dispersion has been attributed to cloud-cloud
scattering \citep{g91} as well as to the driving effects of
spiral structure \citep{t91}. Since both these models were two-dimensional,
they can give us no guidance on the relative strength of the vertical
and in-plane velocity dispersions.

In contrast, there have been a number of studies of the stellar
velocity ellipsoid \citep{d65,db98,e06}. If the velocity
ellipsoid of any stars can give us a clue to that of GMCs, it will be
the youngest O and B type stars, which may be sufficiently young to
still trace the motions of their parent clouds. The smallest values
for the velocity ellipsoid can be found in \citet{d65}, who
measured a relative velocity dispersion $\sigma_z/\sigma_r = 0.75$ for
35 O-B5 supergiants. \citet{db98} measure a significantly smaller
value of 
$\sigma_z/\sigma_r = 0.38^{+0.5}_{-0.10}$ for 500 main sequence stars with
$-0.24 < B-V < 0.14$. However, this colour range extends well into the
A star range where main sequence lifetimes are larger than 100
Myr. Since this velocity dispersion ratio is well known to decrease
with the age of the stellar population \citep{d65,db98}, this value
may not be an appropriate one to adopt for GMCs. 
In a recent study
of OB stars using Hipparcos data, \citet{e06} obtained a value of
$\sigma_z/\sigma_r = 0.64\pm 0.06$ for 800 stars with spectral types
O-B6 within 1 kpc of the Sun. Dividing the sample into stars in the
Gould's Belt and the Local 
Galactic Disk results in values of 
$\sigma_z/\sigma_r = 0.72\pm 0.08$
and $\sigma_z/\sigma_r = 0.57\pm 0.09$, respectively.

\citet{g91} caution that a critical difference
between star-cloud and cloud-cloud scattering, namely the relative
sizes of the epicyclic amplitude and the cloud tidal radius, prevents
direct application of stellar results to molecular clouds. However,
this model ignores the possible effect of spiral arms, which could
conceivably introduce an anisotropy into the cloud motions.
In estimating the possible correction to our observed velocity
dispersions for the 
effects of anisotropy in the velocity ellipsoid, we adopt a
conservative value of $\sigma_z/\sigma_r = 0.6$, which is the
1$\sigma$ upper limit obtained by \citet{e06} for their entire sample.
With this value, corrections for the velocity ellipsoid only exceed
10\% for inclinations greater than 27$^o$.
\label{lastpage}


\begin{thebibliography}{}
\bibitem[Balbus \& Hawley(1991)]{bh91}Balbus, S. A. \& Hawley, J. F. 1991, ApJ, 376, 214
\bibitem[Balbus \& Hawley(1998)]{bh98}Balbus, S. A. \& Hawley, J. F. 1998, Rev. Mod. Phys. 70, 1
\bibitem[Berry et al.(2007)]{b07} Berry, D. S., Reinhold, K., Jenness,
  T., \& Economou, F., 2007, in Astronomical Data Analysis Software
  and Systems XVI, ASP Conference Series, Vol. 376, eds. R. 
A. Shaw, F. Hill \& D. J. Bell., 425
\bibitem[Bigiel et al.(2009)]{b09}Bigiel, F., Leroy, A. K., Walter, F., Brinks,
  E.,  de Blok, W. J. G., Madore, B., \& Thornley, M. D.,  2008,
AJ, 136, 2846
\bibitem[Brunt(2003)]{b03}Brunt, C. 2003, ApJ, 583, 280
\bibitem[Buckle et al.(2009)]{bu09} Buckle, J. V., et al., 2009,
MNRAS, 399, 1026
\bibitem[Buta et al.(2007)]{buta07} Buta, R. J., Corwin, H. G.,
  Odewahn, S. C. 2007, The 
de Vaucouleurs Atlas of Galaxies (Cambridge: Cambridge University Press)
\bibitem[Calzetti et al.(2007)]{c07}Calzetti, D. et al. 2007, ApJ,  666, 870
\bibitem[Cayatte et al.(1990)]{c90} Cayatte,V., van Gorkom, J. H.,
  Balkowski, C., \& Kotanyi, C.,   1990, AJ, 100, 604
\bibitem[Clemens(1985)]{c85}Clemens, D. P., 1985, APJ, 295, 422
\bibitem[Combes \& Becquaert(1997)]{cb97}Combes, F. \& Becquaert,
  J.-F., 1997, A\&A, 327, 453
\bibitem[Coppin et al.(2008)]{copp08}Coppin, K. E. K., et al. 2008,
  MNRAS, 389, 45
\bibitem[Currie et al.(2008)]{c08}	
Currie, M. J., Draper, P. W., Berry, D. S., Jenness, T., Cavanagh, B.,
\& Economou, F., 
2008, in Astronomical Data Analysis Software
  and Systems, ASP Conference Series, Vol. 394, eds. 
R. W. Argyle, P. S. Bunclark, \& J. R. Lewis, 650
\bibitem[Daddi et al.(2008)]{d08} Daddi, E., Dannerbauer, H., 
  Elbaz, D., Dickinson, M., Morrison, G. E., Stern, D., \&
  Ravindranath, S., 2008, ApJ, 673, L21 
\bibitem[Dannerbauer et al.(2009)]{d09}Dannerbauer, H., Daddi, E.,
  Riechers, D. A., Walter, F., Carilli, C. L., Dickinson, M., Elbaz,
  D., \& Morrison, G. E. 2009, ApJ, 698, L178 
\bibitem[Daddi et al.(2010)]{d10} Daddi, E., %Bournaud, F., Walter, F.,
  et al., 2010, ApJ, 714, 118
\bibitem[de Blok \& Walter(2006)]{d06}de Blok, W. J. G. \& Walter,
  F. 2006, AJ, 131, 363  
\bibitem[de Blok et al.(2008)]{dB09} de Blok, W. J. G., Walter, F.,
  Brinks, E., Trachternach, C., Oh, S.-H., \& Kennicutt, R. C., 2008, AJ,
  136, 2648
\bibitem[Delhaye(1965)]{d65}Delhaye, J., 1965, in Galactic Structure,
  Eds. A. Blaauw \& M. Schmidt, University of Chicago Press, 61
\bibitem[Dehnen \& Binney(1998)]{db98}Dehnen, W., \& Binney, J., 1998,
  MNRAS, 298, 387
\bibitem[Dickey et al.(1990)]{d90}Dickey, J. M., Hanson, M. M., \& Helou, G. 1990, ApJ, 352, 522 
\bibitem[Dickey(2009a)]{d09a}Dickey, J. 2009a, arXiv:0901.2380v1
%\bibitem[Dickey(2009b)]{d09b}Dickey, J. 2009b, arXiv:0901.2379v1
\bibitem[Dickman et al.(1985)]{dss85} Dickman, R. L., Snell, R. \&
  Schloerb, F. P. 1985, ApJ, 309, 326
%\bibitem[Faber et al.(2007)]{f07}Faber, S. M., et al., 2007, ApJ, 665, 265
\bibitem[Freedman et al.(2001)]{f01} Freedman, W. L., et al., 2001,
  ApJ, 553, 47
\bibitem[Elias et al.(2006)]{e06}Elias, F., Alfaro, E. J., \&
  Cabrera-Ca\~no, J., 2006, AJ, 132, 1052
\bibitem[Gammie, Ostriker \& Jog(1991)]{g91}Gammie, C. F., Ostriker,
  J. P.,  \& Jog, C. J.,  1991, ApJ, 378, 565
\bibitem[Garc\'{\i}a-Burillo et al.(1993)]{gb93} Garc\'{\i}a-Burillo,
  S., Combes, F., \& Gerin, M., 1993, A\&A, 274, 148
\bibitem[Hanasz \& Lesch(2000)]{hl00}Hanasz, M., \& Lesch, H. 2000, ApJ, 543, 235
%\bibitem[Hanasz \& Lesch(2004)]{hl04}Hanasz, M., \& Lesch, H. 2004, Space Science Reviews, 99, 231
\bibitem[Heiles \& Troland(2003)]{ht03}Heiles, C., \& Troland, T. H. 2003, ApJ, 586, 1067
\bibitem[Henriksen \& Turner(1984)]{h84}Henriksen, R. N.
  \& Turner, B. E. 1984, ApJ, 287, 200
%\bibitem[Ishizuki \& Scoville(1999)]{i99}  Ishizuki, S., \& Scoville, N. Z., 1999, ApJS, 124, 403
\bibitem[Jarrett et al.(2003)]{j03}Jarrett, T. H., Chester, T.,
  Cutri, R., Schneider, S. E.  \& Huchra, J. P.,
 2003, AJ, 125, 525
\bibitem[Jog \& Ostriker(1988)]{jo88}Jog, C. J., \& Ostriker, J. P.,
  1988, ApJ, 328, 404
\bibitem[Jog \& Solomon(1984a)]{j84a}Jog, C. J., \& Solomon, P. M. ,
  1984, ApJ, 276, 114
\bibitem[Jog \& Solomon(1984b)]{j84b}Jog, C. J., \& Solomon, P. M. ,
  1984, ApJ, 276, 127
%\bibitem[]{}Kalberla, P. M. W., \& Dedes, L. 2008, A\&A, 487, 951
\bibitem[Karachentsev et al.(2004)]{k04}Karachentsev, I. D.,
Karachentseva, V. E., Huchtmeier, W. K., \& Makarove, D. I., 2004, AJ,
127, 2031
%\bibitem[Kassin et al.(2007)]{k07}Kassin, S. A., et al., 2007, ApJ,
%  660, L35
\bibitem[Kenney \& Young(1989)]{ky89} Kenney, J. D. P. \& Young,
  J. S., 1989, ApJ, 344, 171
\bibitem[Kennicutt(1989)]{k89}Kennicutt, R. C., 1989, ApJ, 344, 685
\bibitem[Kennicutt(1998)]{k98}Kennicutt, R. C. 1998, ARA\&A, 36, 189
\bibitem[Kennicutt et al.(2003)]{k03}Kennicutt, R. C. et al. 2003,
  PASP, 115, 928 
\bibitem[Kennicutt et al.(2007)]{kenn07}Kennicutt, R. C. et al. 2007,
ApJ, 671, 333
\bibitem[Knapen et al.(1993)]{k93} Knapen, J. H., Cepa, J., Beckman,
  J. E., del Rio, M. S., \& Pedlar, A., 1993, ApJ, 416, 463
\bibitem[Koopman et al.(2001)]{k01} Koopman, R. A., Kenney,
  J. D. P., \& Young, J. S. 2001, ApJS, 135, 125
\bibitem[Kranz et al.(2003)]{kranz03}Kranz, T., Slyz, A., \&
  Hans-Walter Rix, H.-W., 2003, ApJ, 586, 143
\bibitem[Larson et al.(1980)]{l80}Larson, R. B., Tinsley, B. M., \& Caldwell,
C. N. 1980, ApJ, 237, 692
\bibitem[Leonard et al.(2002)]{l02} Leonard, D. C., et al., 2002, AJ,
  124, 2490
\bibitem[Leroy et al.(2008)]{l08} Leroy, A. K., Walter, F., Brinks,
  E., Bigiel, F., de Blok, W. J. G., Madore, B., \& Thornley, M. D.,  2009,
  AJ, 136, 2782
\bibitem[Lockwood(2001)]{l01}Lockwood, F. J. 2001, ApJ, 580, L47
\bibitem[Malhotra(1995)]{m95}Malhotra, S. 1995, ApJ, 448, 138
\bibitem[Maloney \& Black(1988)]{mb88} Maloney, P. \& Black, J. 1988,
  ApJ, 325, 389
\bibitem[McKee(1989)]{m89}McKee, C. F., 1989, ApJ, 345, 782
\bibitem[Mei et al.(2007)]{m07} Mei, S., et al. 2007, ApJ, 655, 144
\bibitem[Mould et al.(2000)]{m00} Mould, J. R., et al. 2000, ApJ, 529, 786
\bibitem[Parker(1966)]{p66}Parker, E. N. 1966, ApJ, 145, 811
\bibitem[Paturel et al.(2000)]{p00} Paturel, G., Fang, Y., Garnier,
  R., Petit, C., \& Rousseau, J.,  2000, A\&AS, 146, 19
\bibitem[Petric \& Rupen(2007)]{pr07}Petric, A. O., \& Rupen, M. P. 2007, AJ, 134, 1952
\bibitem[Phookun et al.(1993)]{p93} Phookun, B., Vogel, S. N., \&
  Mundy, L. G., 1993, ApJ, 418, 113
\bibitem[Piontek \& Ostriker(2007)]{po07} Piontek, R. A., \& Ostriker,
  E. C., 2007, ApJ, 663, 183
\bibitem[Rafikov(2001)]{r01}Rafikov, R. R., 2001, MNRAS, 323, 445
\bibitem[Sanders et al.(1985)]{sss85}Sanders, D. B., Scoville, N. Z.,
  \& Solomon, P. M., 1985, ApJ, 289, 373
\bibitem[Sanders et al.(2003)]{s03}Sanders, D. B., Mazzarella, J. M.,
  Kim, D.-C., Surace, J. A., \& Soifer, B. T., 2003, AJ, 126, 1607
\bibitem[Sellwood \& Balbus(1999)]{sb99}Sellwood, J. A., \& Balbus,
  S. A. 1999, ApJ, 511, 660 
\bibitem[Solomon et al.(1987)]{s87}Solomon, P. M., Rivolo, A. R.,
  Barrett, J., \& Yahil, A., 1987, ApJ, 319, 730
\bibitem[Stark(1984)]{s84}Stark, A. A., 1984, ApJ, 281, 624
\bibitem[Stark \& Brand(1989)]{sb89}Stark, A. A., \& Brand, J., 1989,
  ApJ, 339, 763
\bibitem[Stark \& Lee(2005)]{sl05}Stark, A. A., \& Lee, Y., 2005, ApJ,
  619, L159
\bibitem[Stark \& Lee(2006)]{sl06}Stark, A. A., \& Lee, Y., 2006, ApJ,
  641, L113
\bibitem[Stil et al.(2006)]{s06}Stil, J. M., et al. 2006, ApJ, 637, 366
\bibitem[Strong et al.(1988)]{s88} Strong, A. W., et al. 1988, A\&A, 207, 1 
\bibitem[Tacconi et al.(2008)]{t08}Tacconi, L. J., et al. 2008, ApJ,
  680, 246
\bibitem[Tacconi et al.(2010)]{tacc10}Tacconi, L. J., et al., 2010,
Nature, 463, 781
\bibitem[Tamburro et al.(2008)]{tam08}Tamburro, D.; Rix, H.-W.,
  Walter, F., Brinks, E., de Blok, W. J. G., Kennicutt, R. C., \&
  MacLow, M.-M., 2008, AJ, 136, 2872 
\bibitem[Tamburro et al.(2009)]{t09}Tamburro, D., Rix, H.-W., Leroy,
  A. K., Low, M.-M. Mac, Walter, F., Kennicutt, R. C., Brinks, E., \& de
  Blok, W. J., 2009, AJ, 137, 4424
\bibitem[Tasker \& Tan(2009)]{tt09}Tasker, E. J., \& Tan, J. C., 2009,
  ApJ, 700, 358
\bibitem[Thomasson et al.(1991)]{t91}Thomasson, M., Donner, K. J., \&
  Elmegreen, B. G., 1991, A\&A, 250, 316
\bibitem[Tonry et al.(2001)]{t01}Tonry, J. L, Dressler, A.,
  Blakeslee, J. P., Ajhar, E. A., Fletcher, A. B., Luppino,
  G. A., Metzger, M. R., \& Moore, C. B.,  2001, ApJ, 546, 681
\bibitem[Toomre(1964)]{t64}Toomre, A., 1964, ApJ, 139, 1217
\bibitem[van der Kruit \& Shostak(1982)]{sk82} van
  der Kruit, P. C. \&  Shostak, G. S.,1982, A\&A, 105, 351
\bibitem[van Zee \& Bryant(1999)]{zb99}van Zee, L., \& Bryant,
  J. 1999, AJ, 118, 2172 
\bibitem[Vollmer et al.(2008)]{v08} Vollmer, B., Soida, M., Chung, A.,
  van Gorkum, J. H., Otmianowska-Mazur, K., Beck, R.,
Urbanik, M., and Kenney, J. D. P., 2008, A\&A, 483, 89
\bibitem[Walsh et al.(2002)]{w02}Walsh, W., Beck, R., Thuma, G.,
  Weiss, A., Wielebinski, R., \& Dumke, M., 2002, A\&A, 388, 7
\bibitem[Walter et al.(2008)]{w08}Walter, F., Brinks,
  E.,  de Blok, W. J. G., Bigiel, F., Kennicutt, R. C., Thornley,
  M. D., \& Leroy, A. K.,  2008, AJ, 136, 2563
\bibitem[Warren et al.(2010)]{w10}Warren, B. E., et
  al., 2010, ApJ, 714, 571
\bibitem[Williams et al.(1994)]{w94}Williams, J. P., de Geus, E. J.,
  \& Blitz, L.  1994, ApJ, 428, 693
\bibitem[Wilson et al.(2009)]{w09}Wilson, C. D., et al., 2009,
  ApJ, 693, 1736
\bibitem[Wilson \& Scoville(1990)]{ws90} Wilson, C. D. \& Scoville,
  N. Z., 1990, ApJ, 363, 435
\bibitem[Yang et al.(2007)]{y07} Yang, C.-C., Gruendl, R. A., Chu,
  Y.-H., Mac Low, M., \& Fukui, Y., 2007, ApJ, 671, 374
\bibitem[Young \& Scoville(1991)]{ys91} Young, J. S. \& Scoville,
  N. Z. 1991, ARAA, 29, 581
\bibitem[Young et al.(1995)]{y95} Young, J. S., et al., 1995, ApJS,
  98, 219
\bibitem[Young \& Lo(1997)]{l97}Young, L. M. \& Lo, K. Y., 1997, ApJ
  490 710 
\bibitem[Young \& Lo(1996)]{l96}Young, L. M. \& Lo, K. Y., 1996, ApJ,
  462, 203   
\end{thebibliography}
\end{document}